\documentclass[twocolumn,showpacs,preprintnumbers,amsmath,amssymb]{revtex4}

\usepackage{graphicx}
\usepackage{dcolumn}
\usepackage{bm}

\begin{document}

\title{Randomizing world trade. II. A weighted network analysis}

\author{Tiziano Squartini}
\affiliation{CSC and Department of Physics, University of Siena, Via Roma 56, 53100 Siena (Italy)}
\affiliation{Lorentz Institute for Theoretical Physics, Leiden Institute of Physics, University of Leiden, Niels Bohrweg 2, 2333 CA Leiden (The Netherlands)}
\author{Giorgio Fagiolo}
\affiliation{LEM, Sant'Anna School of Advanced Studies, 56127 Pisa (Italy)}
\author{Diego Garlaschelli}
\affiliation{Lorentz Institute for Theoretical Physics, Leiden Institute of Physics, University of Leiden, Niels Bohrweg 2, 2333 CA Leiden (The Netherlands)}%

\date{\today}

\begin{abstract}
Based on  the misleading expectation that weighted network properties always offer a more complete description than purely topological ones, current economic models of the International Trade Network (ITN) generally aim at explaining local weighted properties, not local binary ones. Here we complement our analysis of the binary projections of the ITN by considering its weighted representations. We show that, unlike the binary case, all possible weighted representations of the ITN (directed/undirected, aggregated/disaggregated) cannot be traced back to local country-specific properties, which are therefore of limited informativeness. Our two papers show that traditional macroeconomic approaches systematically fail to capture the key properties of the ITN. In the binary case, they do not focus on the degree sequence and hence cannot characterize or replicate higher-order properties. In the weighted case, they generally focus on the strength sequence, but the knowledge of the latter is not enough in order to understand or reproduce indirect effects. 
\end{abstract}



\maketitle

\section{Introduction}
In this paper we extend our analysis of the binary projection of the International Trade Network (ITN) reported in the previous paper \cite{part1} to the weighted representation of the same network.
As in the binary case, we employ a recently-proposed randomization method \cite{myrandomization} to assess in detail the role that local properties have in shaping higher-order patterns of the weighted ITN in all its possible representations (directed/undirected, aggregated/disaggregated) and across several years. In the weighted case, we employ a null model that preserves on average the strengths of the vertices only. More specifically, when the network is undirected, node strength is preserved on average, whereas when the network is directed in- and out-strengths are conserved  separately on average. From a trade perspective, this means that the null model controls for country total trade in the undirected case, and for country total imports and exports (as a share of total world yearly trade) in the directed case. This implies that degrees are not preserved on average in either case. For example, in the undirected case, a country preserves on average (over all graphs accounted for by the null model) its total observed trade flow, but not its observed number of partners. Preserving total trade and number of partners simultaneously for each country is a more severe constraint which goes beyond the scope of this paper, since it would not allow us to compare our results with well established international-economics approaches which only take total trade, and not the number of partners, into account.
 
We find that, unlike the binary case, higher-order patterns of weighted (either directed or undirected, either aggregated or disaggregated) representations of the ITN cannot be merely traced back to local properties alone (i.e., node strength sequences). In particular, when compared to its randomized variants, the observed weighted ITN displays a different and sparser topology (despite the ITN is usually considered denser than most studied networks), stronger disassortativity, and larger clustering. As sparser and less aggregated commodity-specific representations are considered, the accordance between the real and randomized networks gets even worse. All these results hold for both undirected and directed projections, and are robust throughout the time interval we consider (from year 1992 to 2002).

From an international-trade perspective, our results indicate that a weighted network description of trade flows, by focusing on higher-order properties in addition to local ones, captures novel and fresh evidence. Therefore, traditional analyses of country trade profiles focusing only on local properties and country-specific statistics convey a partial description of the richness and details of the ITN architecture. 
Moreover, economic models and theories that only aim at explaining the local properties of the weighted ITN (i.e. the total values of imports and exports of world countries) are limited, as such properties  have no predictive power on the rest of the structure of the network.

We refer the reader to the companion paper \cite{part1} for a description of the data, the notation used, the meaning and economic importance of local topological properties, and the randomization method that we have adopted.

\section{The ITN as a weighted undirected network\label{sec_wun}}
The weighted representation of the ITN takes into account the intensity (dollar value) of trade relationships, and can be either directed or undirected.
The structure of the network is completely specified by the weight matrix $\mathbf{W}$, whose entries $\{w_{ij}\}$ have been defined in Ref. \cite{part1} in the directed and undirected case. 
In the weighted undirected case, an edge between vertices $i$ and $j$ represents the presence of at
least one of the two possible trade relationships between the two countries $i$ and $j$, and the weight $w_{ij}$
represents the average trade value (or equivalently half the total bilateral trade value) \cite{part1}.
Clearly, if no trade occurs in either direction, then $w_{ij}=0$ and no link exists. The weight matrix
$\mathbf{W}$ is therefore symmetric: $w_{ij}=w_{ji}$.  
One can still define an adjacency matrix $\mathbf{A}$, describing the purely binary topology of the network, with entries $a_{ij}=\Theta(w_{ij})$ where $\Theta(x)=1$ if $x>0$ and $\Theta(x)=0$ otherwise. Clearly, the symmetry of $\mathbf{W}$ implies the symmetry of $\mathbf{A}$.

In the case considered, the local constraints $\{C_a\}$ are the strengths of all vertices, i.e.
the \emph{strength sequence} $\{s_i\}$ \cite{part1}.
The randomization method we adopted \cite{myrandomization} proceeds in this case by specifying the constraints $\{C_a\}\equiv \{s_i\}$
(see Appendix \ref{app_wun}), and yields the ensemble probability of any weighted graph $\mathbf
{G}$, which now is uniquely specified by its generic weight matrix $\mathbf{W}$.
For any weighted topological property $X$, it is therefore possible to easily obtain the expectation value
$\langle X\rangle$ across the ensemble of weighted undirected graphs with specified strength sequence.
By construction, the expected strength $\langle s_i\rangle$ across the randomized ensemble is equal to the empirical value
$s_i$, therefore in the weighted undirected case the strength values $\{s_i\}$ are the natural independent variables in terms of which other weighted properties $X$ can be
visualized. By contrast, other properties such as the degree of vertices, and consequently the total number of links, are not preserved on average.

In our analyis, we first use the matrix $\mathbf{W}$, and the strength sequence $\{s_i\}$ obtained from it, as the starting point for the randomization method. 
However, as we mentioned in the companion paper \cite{part1}, in order to allow a consistent temporal analysis we need to focus on the rescaled weights $\tilde{w}_{ij}\equiv w_{ij}/w_{tot}$, where where $w_{tot}=\sum_i\sum_{j<i}w_{ij}$ is the total yearly weight. Consistently, we define the rescaled strength
\begin{equation}
\tilde{s}_i\equiv\sum_{j\ne i}\tilde{w}_{ij}=\frac{s_i}{w_{tot}}
\end{equation}
and we similarly use $\tilde{w}_{ij}$ instead of $w_{ij}$ in the definition of all other weighted topological quantities. This procedure allows for homogeneous comparisons between real and randomized webs, and across different years and commodities. In particular, it filters out the increase of world trade in nominal terms. 
Note that, across the randomized ensemble, $w_{tot}$ is a random variable, since so are the weights $w_{ij}$. However, we can rewrite $w_{tot}=\sum_i s_i/2$, and since $\langle s_i\rangle=s_i$ we have
\begin{equation}
\langle w_{tot}\rangle=\frac{\sum_i \langle s_i\rangle}{2}=\frac{\sum_i s_i}{2}=w_{tot}
\label{eq_wtotund}
\end{equation}
The above result shows that the expectation value of the total weight across the randomized ensemble is constrained by the method to be strictly equal to the observed value $w_{tot}$. In other words, the constraint on the strengths is automatically reflected also in a constraint on the total weight, and we can therefore use the the latter to rescale all weights, both in the real network and in its randomized variants.

In economic terms, specifying the strength sequence amounts to investigate the properties of the trade network once total trade of all countries is controlled for.
It must be noticed that, by controlling for the strength of vertices, one automatically takes at least partially on board considerations related to country-size effects. Indeed, it is common wisdom that bilateral trade flows, and therefore weighted network statistics, should depend on (at the very least) country-specific variables like country gross-domestic product (GDP) and some additional pairwise factors like geographical distance \cite{GravityBook,Fagiolo2010jeic}. Since total trade (and total imports or exports) are known to be positively and strongly correlated with country GDP, our analysis already controls for some size effect. As we discuss in the concluding remarks, a further exercise to be carried out may concern understanding the extent to which other determinants of trade (like distance) may explain the observed properties of the ITN, i.e. building more economically-meaningful models of trade that start explaining ITN properties rather than statistically reproducing them only.

\subsection{Edge weights\label{sec_wun_w}}
We start with the analysis of the completely aggregated network (i.e. $c=0$ according to our notation described in Ref. \cite{part1}).
Therefore, in the following formulas, we set $\mathbf{W}\equiv\mathbf{W}^0$ and drop the superscript for brevity.
Our aim is to understand how specifying the strength sequence affects higher-order network properties. Therefore we will consider the weighted counterparts of the topological properties we have already studied in the binary case \cite{part1}. However, due to the larger number of degrees of freedom, in the weighted case there are also additional quantities to study which have no binary analogue. In particular, it is important to understand the effect that the enforcement of local constraints (strength sequence) has on the weights of the network, as well as on its purely binary topology. It is useful to perform this analysis as a preliminary step, before discussing other results.  

To this end, we start by comparing the empirical weight distribution with the expected one. Importantly, one should not confuse the expected weight distribution with the distribution of expected weights.
In the spirit of our analysis, the empirical network (and so its weight distribution) is regarded as a particular possible realization of the null model with given strengths, and the comparison with the expected properties aims at assessing how unlikely that particular realization is.
Therefore the observed number of edges with weight equal to $w$ (i.e. the empirical weight distribution) should be compared with the expected number of such edges in a single realization (the expected weight distribution), rather than with the number of edges whose expected weight across realizations is equal to $w$ (the distribution of expected weights).
The difference between the two expected quantities is evidenced by the fact that the expected edge weight between vertices is always positive (see Appendix \ref{app_wun}), whereas in a single realization there are a number of zero-weight edges (i.e. missing links).
\begin{figure}[t]
\includegraphics[width=.45\textwidth]{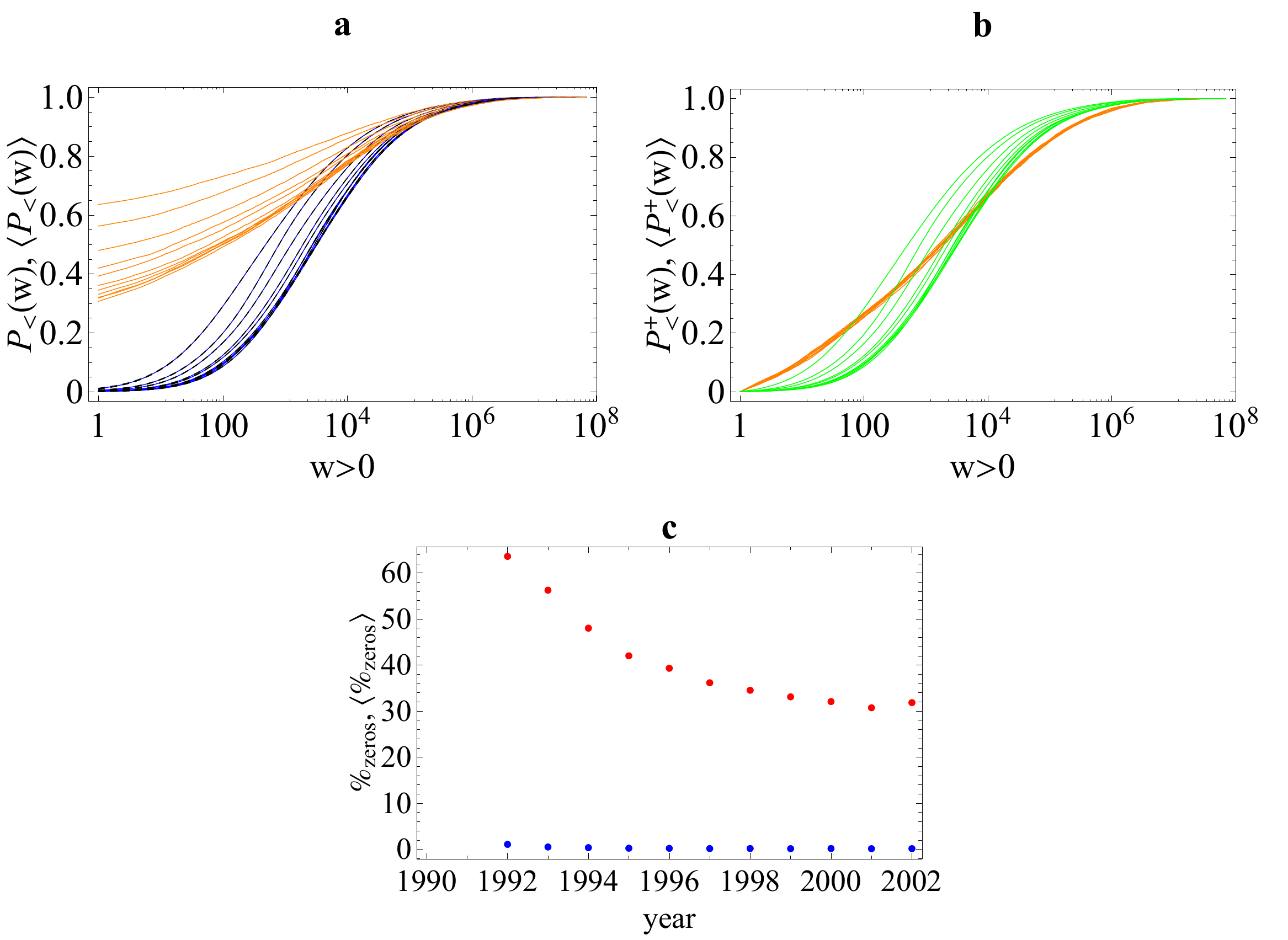}
\caption{\label{fig_wun_w}
(Color online) Edge weights in the weighted undirected ITN. 
\textbf{a)} cumulative distributions of edge weights, for all years from 1992 (top) to 2002 (bottom). Orange (upper curves): real network; blue (lower curves): expectation for the maximum-entropy null model with specified strengths.
\textbf{b)} same as the previous panel, but excluding zero weights (missing links). Orange (dark grey): real network; green (light grey): null model. \textbf{c)} percentage of missing links as a function of time. Red (upper points): real network; blue (lower points): null model.}
\end{figure}

In Fig.~\ref{fig_wun_w}a we therefore compare the cumulative distribution of observed weights $P_<(w)$ (the fraction of edge weights smaller than $w$) with the expected number $\langle P_<(w)\rangle$ (see Appendix \ref{app_wun}), both including missing links ($w=0$) and therefore normalized to the number of pairs of vertices. As an alternative, in Fig.~\ref{fig_wun_w}b we also compare the cumulative distribution of positive weights $P^+_<(w)$ (which excludes missing links and is therefore normalized to the total number of links) with the expected number $\langle P^+_<(w)\rangle$ (see Appendix \ref{app_wun}). We find that, for all years in our time window, the real distributions are always different from the expected ones. To rigorously confirm this, we have performed Kolmogorov-Smirnov and Lilliefors tests and for all years we always had to reject the hypothesis that real and expected distributions are the same ($5\%$ significance level). For the positive weight distributions  $P^+_<(w)$ and $\langle P^+_<(w)\rangle$ we also separately tested the hypothesis of the log-normality of the distributions, and again we always had to reject it  ($5\%$ significance level).

The above results, besides highlighting large differences in the weighted structure of real and randomized networks, also convey important information about remarkable deviations in their topology. The largest difference between the curves $P_<(w)$ and $\langle P_<(w)\rangle$ is found at $w=0$, and the corresponding points $P_<(0)$ and $\langle P_<(0)\rangle$ represent the fractions $\%_{zeros}$ and $\langle\%_{zeros}\rangle$ of zero weights (missing links) in the network. In Fig.~\ref{fig_wun_w}c we show the evolution of these fractions over time. We find that the fraction of missing links in the real network decreases in time over the time interval considered (i.e. the link density increases), but its value is always much larger than the corresponding expected value. 
Thus, despite it is usually considered a very dense graph, with more links per node than most other real-world networks, we find that the ITN turns out to be surprisingly sparser than random weighted networks with the same strength sequence. This fixes a previously unavailable benchmark for the density of the empirical ITN, and implies that the high percentage of missing trade relations among world countries is not explained by size effects (i.e. the total trade value of all countries). Note that this result would be trivial if we were considering trade magnitudes as real-valued, rather than integer-valued, weights. For real-valued weights, the volume of the configuration space (the number of networks in the ensemble) would be infinite, and the probability of networks with zero-valued weights would be zero. So, topologically, almost all networks in the random ensemble would be complete graphs, making every real network sparser than (or at most as sparse as) its randomized counterpart. Instead, in our analysis we consider integer-valued weights (as discussed in detail when presenting our data and methods in the companion paper \cite{part1}) reflecting more correctly the fact that money has indivisible units. This always gives a positive probability $q_{ij}(0)>0$ (see Appendix) to missing links, and as a result the expected density of links is always smaller than one.

In what follows, we study the effects of the strength sequence on higher-order topological properties, in analogy with our binary analysis \cite{part1}. We first report detailed results for the 2002 snapshot of the commodity-aggregated network (Sections \ref{sec_wun_anns} and \ref{sec_wun_clustering}), then consider the temporal evolution of the aggregated network (Section \ref{sec_wun_evolution}), and finally perform a commodity-specific analysis (Section \ref{sec_wun_dis}).

\begin{figure}[t]
\includegraphics[width=.45\textwidth]{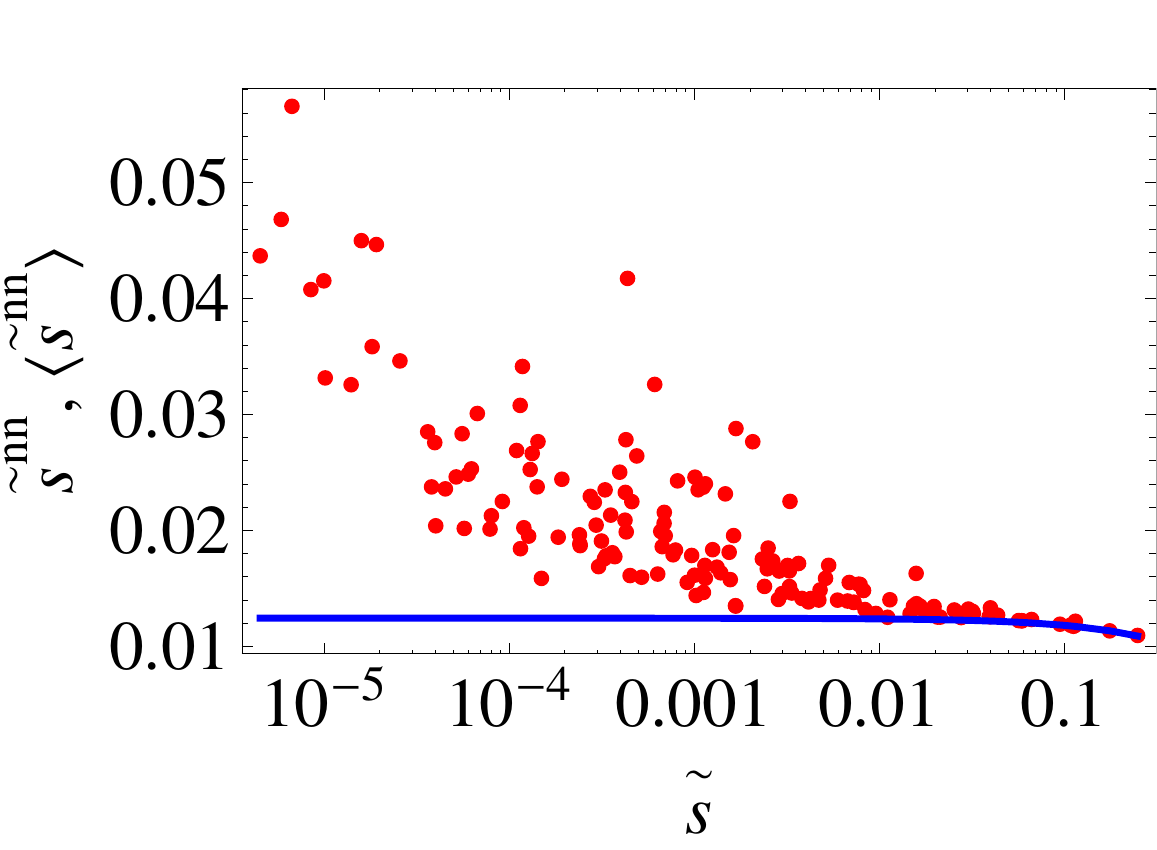}
\caption{\label{fig_wun_snn} (Color online) Average  nearest neighbor strength $\tilde{s}^{nn}_i$ versus strength $\tilde{s}_i$ in the 2002 snapshot of the real weighted undirected ITN (red points), and corresponding average over the maximum-entropy ensemble with specified strengths (blue solid curve).}
\end{figure}
\subsection{Average nearest neighbor strength\label{sec_wun_anns}}
We start with the weighted counterpart of the average nearest neighbor degree (ANND), i.e. the \emph{average nearest neighbor strength} (ANNS) of vertex $i$, defined as
\begin{equation}
\tilde{s}^{nn}_i\equiv\frac{\sum_{j\ne i}a_{ij}\tilde{s}_j}{k_i}
=\frac{\sum_{j\ne i}\sum_{k\ne j}a_{ij}\tilde{w}_{jk}}{\sum_{j\ne i}a_{ij}}
\label{eq_wun_snn}
\end{equation}
The ANNS measures the average strength of the neighbors of a given vertex. Similarly to the ANND, the ANNS involves indirect interactions of length $2$, however (as happens for most weighted quantities) mixing both weighted and purely topological information: in particular, terms of the type $a_{ij}\tilde{w}_{jk}$ appear in the definition.

The correlations between the strength of neighboring countries can be inspected by plotting $\tilde{s}^{nn}_i$ versus $\tilde{s}_i$. This is shown in Fig.~\ref{fig_wun_snn}. Even if the points are now significantly more scattered, we find a decreasing trend as previously observed for the corresponding binary quantities \cite{part1}.
This trend signals that highly trading countries trade typically with poorly trading ones (and vice versa), confirming on a weighted basis the disassortative character observed at the binary level. However, in this case the null model behaves in a completely different way: over the randomized ensemble with specified strength sequence, the expectation value $\langle \tilde{s}^{nn}_i\rangle$ of the ANNS (see Appendix \ref{app_wun}) decreases over a much narrower range (see Fig.~\ref{fig_wun_snn}), and is always different from the observed value. 

This important results implies that, even if we observe disassortativity in both cases (binary and weighted), we find that in the binary case this property is completely explained by the degree sequence, whereas in the weighted case it is not explained by the strength sequence. This has implications for economic models of international trade: while no theoretical explanation is required in order to explain why poorly connected countries trade with highly connected ones on a binary basis (once the number of trade partners is specified), additional explanations are required in order to explain the same phenomenon at a weighted level, even after controlling for the total trade volumes of all countries.
This result could look counterintuitive, as a simple visual inspection would suggest that in the binary case the disassortative behavior is in absolute terms less noisy, and sometimes more pronounced, than in the weighted one.

\begin{figure}[t]
\includegraphics[width=.45\textwidth]{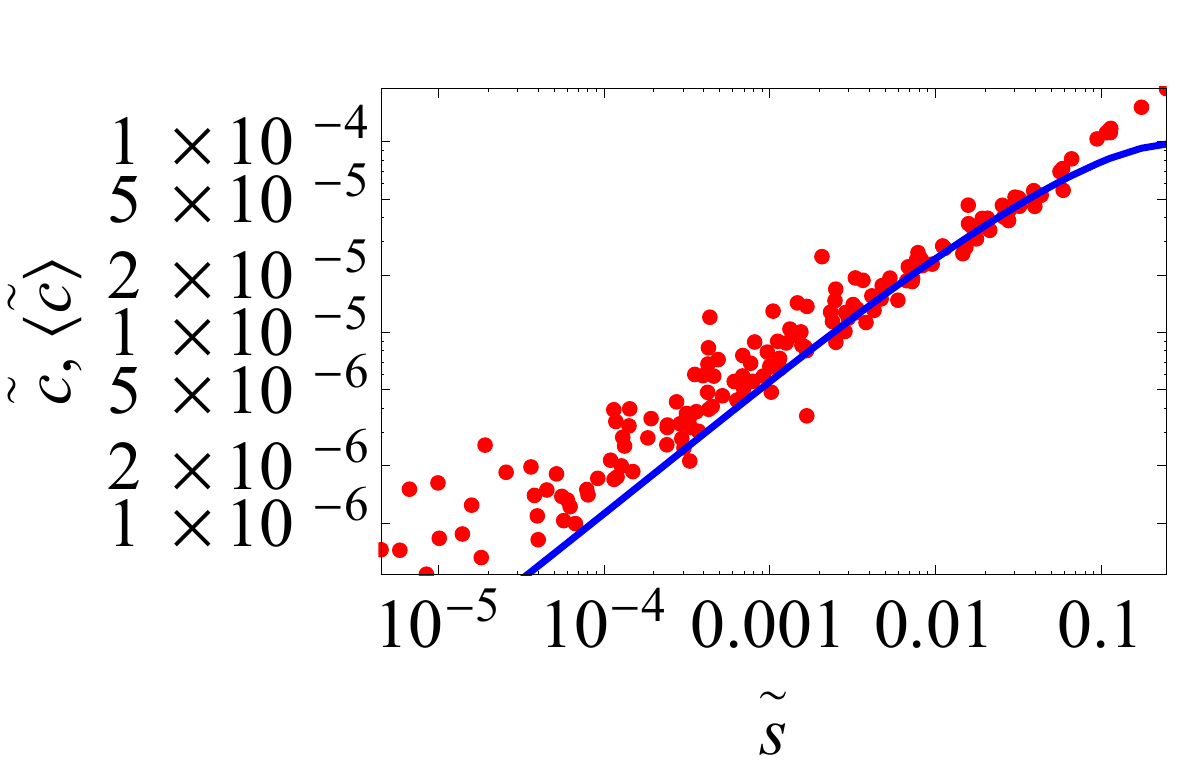}
\caption{\label{fig_wun_cs} (Color online) Weighted clustering coefficient $\tilde{c}_i$ versus strength $\tilde{s}_i$ in the 2002 snapshot of the real weighted undirected ITN (red points), and corresponding average over the maximum-entropy ensemble with specified strengths (blue solid curve).}
\end{figure}

\subsection{Weighted clustering coefficient\label{sec_wun_clustering}}
We now consider the weighted version of
the clustering coefficient. In particular, we
choose the definition proposed in
Ref.~\cite{kertesz_definitionclustering},
which has a more direct extension to the
directed case \cite{Fagiolo2007pre}.
According to that definition, the
(rescaled) \emph{weighted clustering coefficient} $\tilde{c}_i$ represents the
intensity of the triangles in which vertex $i$
participates:
\begin{eqnarray}
\tilde{c}_i&\equiv&\frac{\sum_{j\neq i}
\sum_{k\ne i,j}(\tilde{w}_{ij}\tilde{w}
_{jk}\tilde{w}_{ki})^{1/3}}{k_{i}(k_
{i}-1)}\nonumber\\
&=&\frac{\sum_{j\neq i}\sum_{k\ne i,j}
(\tilde{w}_{ij}\tilde{w}_{jk}\tilde{w}_
{ki})^{1/3}}{\sum_{j\neq i}\sum_
{k\ne i,j}a_{ij}a_{ik}}
\label{eq_wun_c}
\end{eqnarray}
Note that $\tilde{c}_i$ takes into account
indirect interactions of length $3$,
corresponding to products of the type
$\tilde{w}_{ij}\tilde{w}_{jk}\tilde{w}_
{ki}$ appearing in the above formula.
In Fig.~\ref{fig_wun_cs} we plot $\tilde
{c}_i$ versus $\tilde{s}_i$.
This time we find an increasing trend of
$\tilde{c}_i$ as a function of $\tilde{s}
_i$, indicating that countries with larger
total trade participate in more intense
trade triangles. We also show the trend
followed by the randomized quantity
$\langle \tilde{c}_i\rangle$ (see
Appendix \ref{app_wun}), which is found
to approximately reproduce the empirical
data.
Despite the partial accordance between the clustering profile of real and randomized networks, the total level of clustering of the real ITN is however larger than its randomized counterpart, as we show below (Section \ref{sec_wun_evolution}) for all the years considered.

\begin{figure}[t]
\includegraphics[width=.45\textwidth]{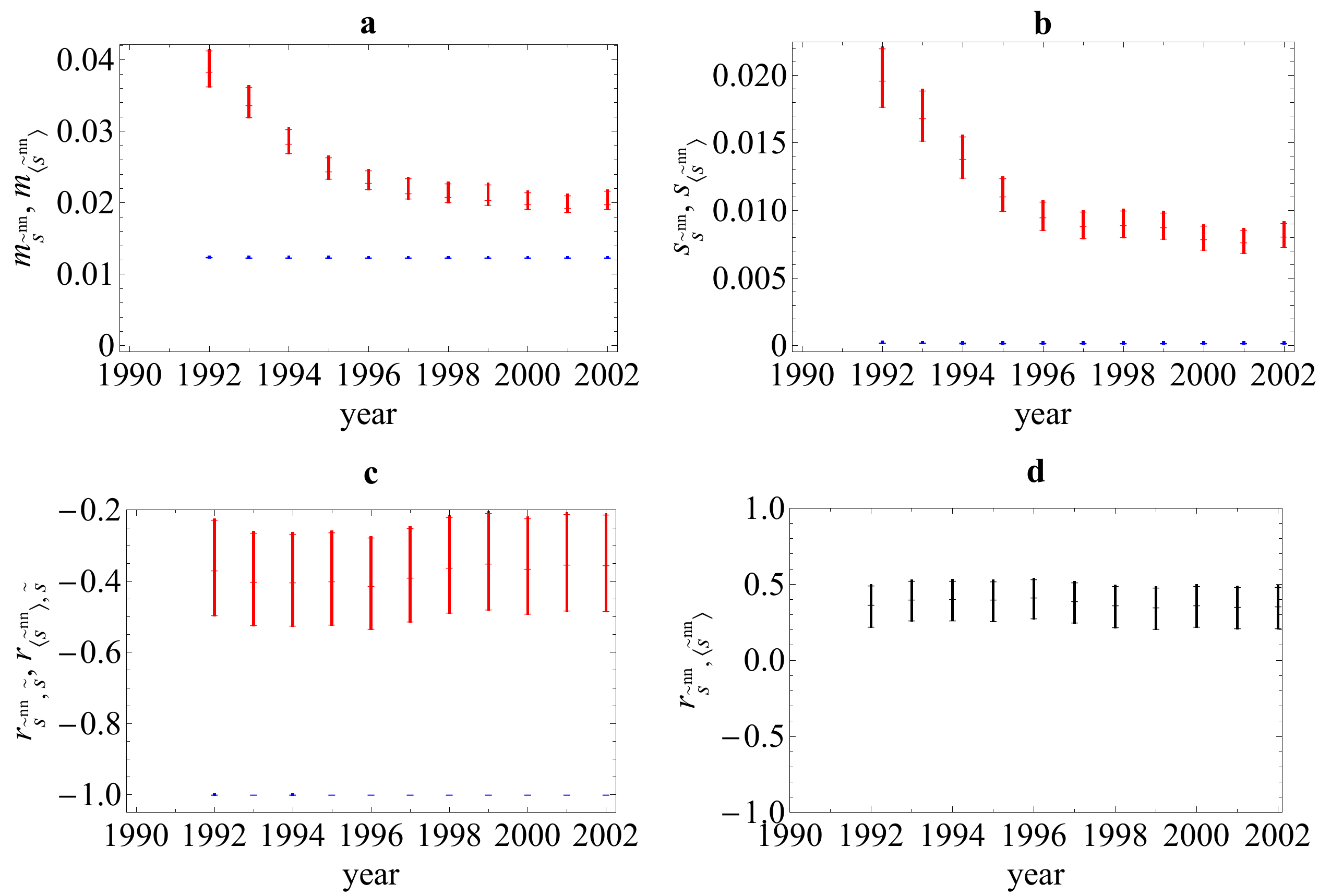}
\caption{\label{fig_wun_snn_t}
(Color online) Temporal evolution of the properties of the (rescaled) average nearest neighbor strength $\tilde{s}^{nn}_i$ in the 1992-2002 snapshots of the real weighted undirected ITN and of the corresponding null model with specified strengths. \textbf{a)} average of $\tilde{s}^{nn}_i$ across all vertices (red, upper symbols: real data; blue, lower symbols: null model). 
\textbf{b)} standard deviation of $\tilde{s}^{nn}_i$ across all vertices  (red, upper symbols: real data; blue, lower symbols: null model). 
\textbf{c)} correlation coefficient between $\tilde{s}^{nn}_i$ and $\tilde{s}_i$ (red, upper symbols: real data; blue, lower symbols: null model). 
\textbf{d)} correlation coefficient between $\tilde{s}^{nn}_i$ and $\langle \tilde{s}^{nn}_i\rangle$. The $95\%$ confidence intervals of all quantities are represented as vertical bars.}
\end{figure}
\subsection{Evolution of weighted undirected properties\label{sec_wun_evolution}}
The results we have reported above are
qualitatively similar for each of the 11
shapshots of the ITN from year 1992 to
2002. As for our binary analyses \cite{part1}, we can
therefore compactly describe the
temporal evolution of weighted undirected
properties in terms of simple indicators.

We start with the analysis of the ANNS (Fig.~\ref{fig_wun_snn_t}).
In Fig.~\ref{fig_wun_snn_t}a we report the
average (across vertices) and the associated $95\%$ confidence
interval of both real and randomized
values ($\{\tilde{s}^{nn}_i\}$ and
$\{\langle\tilde{s}^{nn}_i\rangle\}$) as a
function of time.
We find that the average of $\tilde{s}^{nn}
_i$ has been first decreasing rapidly, and has then remained almost constant. This behavior is already clean from trends in the total volume of trade, since all weights have been rescaled and divided by $w_{tot}$.
By contrast, the average of the randomized quantity $\langle\tilde{s}^{nn}_i\rangle$ displays a constant trend throughout the time interval considered, and its value is always significantly smaller than the empirical one.
Thus, unlike the binary case, the null model does not reproduce the average values of the correlations considered, and does not capture their temporal evolution.
A similar behavior is observed for the evolution of the standard deviation of the ANNS across vertices (Fig.~\ref{fig_wun_snn_t}b).
In Fig.~\ref{fig_wun_snn_t}c we show the correlation coefficient between the empirical quantities
$\{\tilde{s}^{nn}_i\}$ and $\{\tilde{s}_i\}$, whose value (fluctuating around $-0.4$) compactly summarizes the disassortativity of the noisy scatter plot that we have shown previously in Fig.~\ref{fig_wun_snn}, and the correlation coefficient between the randomized quantities $\{\langle\tilde{s}^{nn}_i\rangle\}$ and $\{\langle \tilde{s}_i\rangle\}=\{\tilde{s}_i\}$, which instead displays a different value close to $-1$ (due to the noise-free, even if much weaker, decrease of the randomized curve in Fig.~\ref{fig_wun_snn}).
The discrepancy between the null model and the real network is finally confirmed by the small correlation between $\{\tilde{s}^{nn}_i\}$ and $\{\langle\tilde{s}^{nn}_i\rangle\}$ (Fig.~\ref{fig_wun_snn_t}d), which is in marked contrast with the perfect correlation between $\{k^{nn}_i\}$ and $\{\langle k^{nn}_i\rangle\}$ we found in the binary case.

\begin{figure}[t]
\includegraphics[width=.45\textwidth]{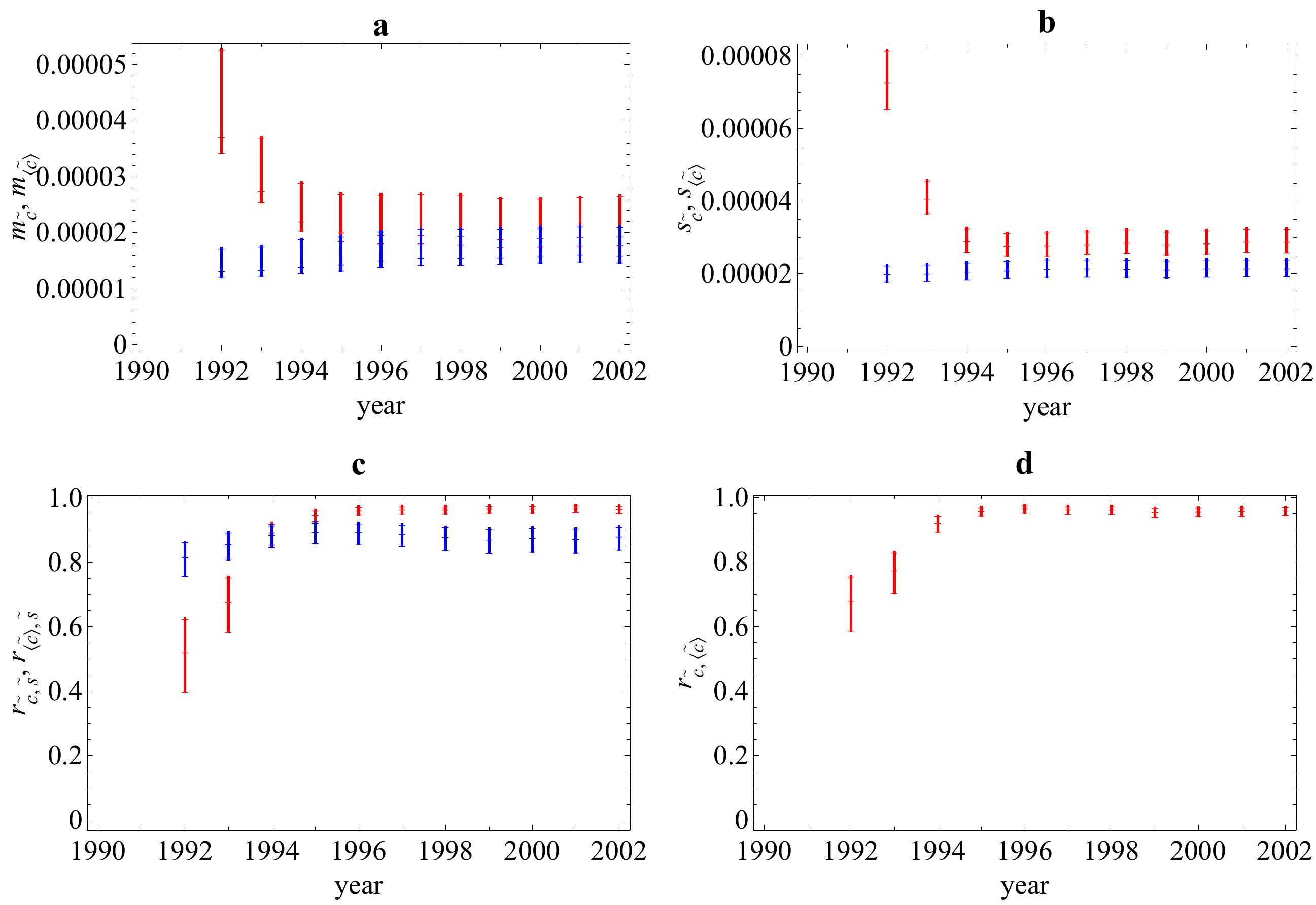}
\caption{\label{fig_wun_c_t}
(Color online) Temporal evolution of the properties of the (rescaled) weighted clustering coefficient $\tilde{c}_i$ in the 1992-2002 snapshots of the real weighted undirected ITN and of the corresponding null model with specified strengths. 
\textbf{a)} average of $\tilde{c}_i$ across all vertices (red, upper symbols: real data; blue, lower symbols: null model). 
\textbf{b)} standard deviation of $\tilde{c}_i$ across all vertices (red, upper symbols: real data; blue, lower symbols: null model). 
\textbf{c)} correlation coefficient between $\tilde{c}_i$ and $\tilde{s}_i$ (red, initially lower symbols: real data; blue, initially upper symbols: null model). 
\textbf{d)} correlation coefficient between $\tilde{c}_i$ and $\langle \tilde{c}_i\rangle$. The $95\%$ confidence intervals of all quantities are represented as vertical bars.}
\end{figure}

In Fig.~\ref{fig_wun_c_t} we report a similar analysis for the evolution of the weighted clustering coefficient. We find that, despite the partial accordance of the real and randomized clustering profiles shown in Fig.~\ref{fig_wun_cs}, the average level of clustering of the real network is always higher than its randomized variant (Fig.~\ref{fig_wun_c_t}a), even if the two values have become closer through time. The same is true for the standard deviation of the weighted clustering coefficient (Fig.~\ref{fig_wun_c_t}b). We also find that the correlation coefficient between the empirical quantities $\{\tilde{c}_i\}$ and $\{\tilde{s}_i\}$ (Fig.~\ref{fig_wun_c_t}c) has rapidly increased between the years 1992 and 1995 (from about $0.5$ to more than $0.95$) and has then remained stable in time. This indicates that the scatter plot shown in Fig.~\ref{fig_wun_cs} for the year 2002 becomes noisier in the first snapshots of our time window, as we confirmed through an explicit inspection (not shown). By contrast, the correlation coefficient between the randomized quantities $\{\langle\tilde{c}_i\rangle\}$ and $\{\langle\tilde{s}_i\rangle\}=\{\tilde{s}_i\}$ displays much smaller variations about the value $0.85$, and is therefore initially larger, and eventually smaller, than the corresponding empirical value.
Finally, in Fig.~\ref{fig_wun_c_t}d we show the correlation coefficient between $\{\tilde{c}_i\}$ and $\{\langle\tilde{c}_i\rangle\}$. The increasing trend confirms the growing agreement between the real and randomized  clustering coefficients, already suggested by the previous plots. Note however that even two perfectly correlated lists of values (correlation coefficient equal to $1$) are only equal if their averages are the same (otherwise they are simply proportional to each other). Thus large correlation coefficients between two quantities can only be interpreted in conjunction with a comparison of the average values of the same quantities. While in the binary case we simultaneously found perfect correlation and equal average values between real and randomized quantities \cite{part1}, in this case we find large correlation but different average values, systematically confirming only a partial accordance between the real network and the null model.

\begin{figure}[t]
\includegraphics[width=.45\textwidth]{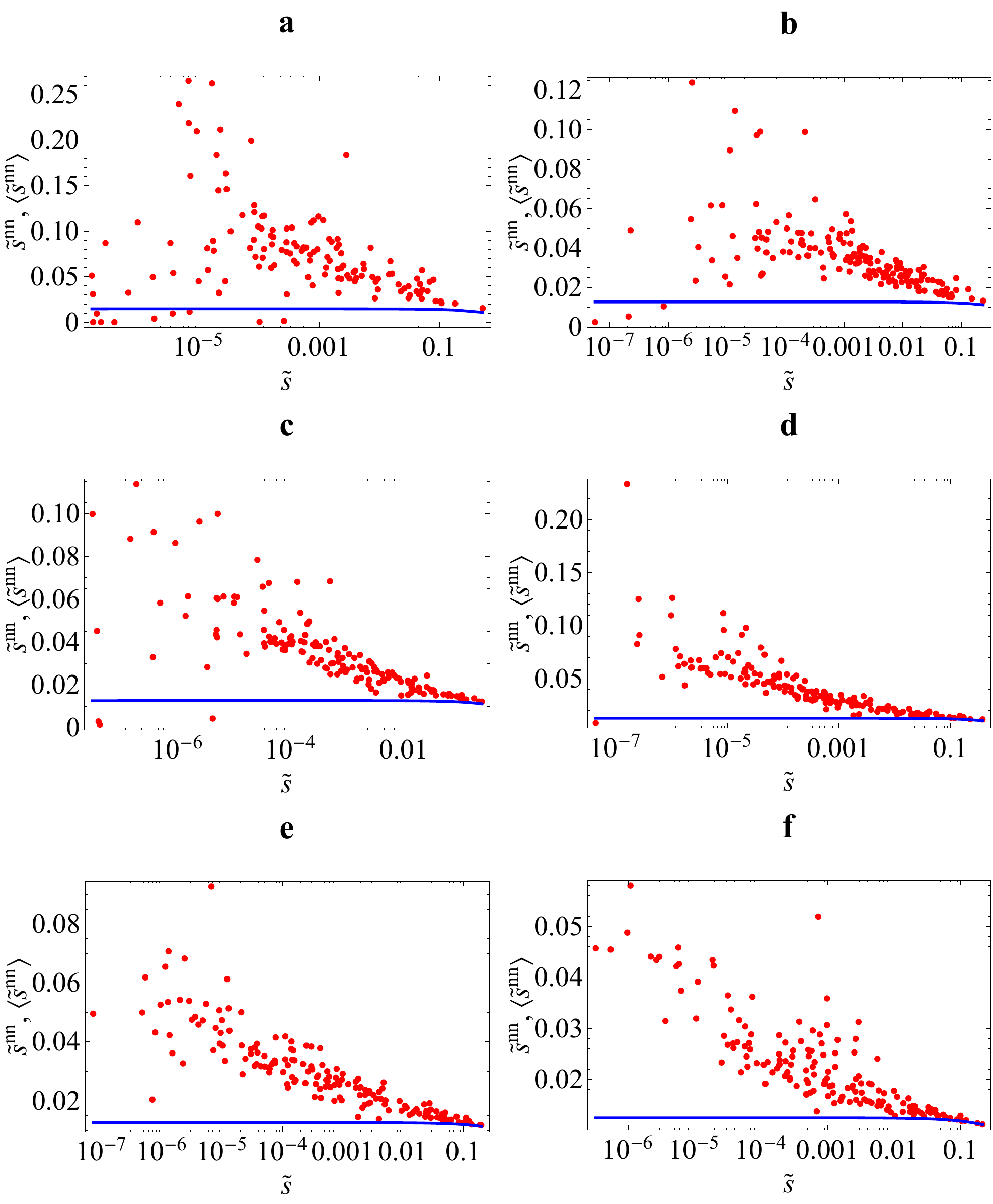}
\caption{\label{fig_wun_dis1}
(Color online) Average nearest neighbor strength $\tilde{s}^{nn}_i$ versus strength $\tilde{s}_i$ in the 2002 snapshots of the commodity-specific (disaggregated) versions of the real weighted undirected ITN (red points), and corresponding average over the maximum-entropy ensemble with specified strengths (blue solid curves).
\textbf{a)} commodity 93; 
\textbf{b)} commodity 09; 
\textbf{c)} commodity 39; 
\textbf{d)} commodity 90; 
\textbf{e)} commodity 84; 
\textbf{f)} aggregation of the top 14 commodities (see Ref. \cite{part1} for details). From a) to f), the intensity of trade and level of aggregation increases.} 
\end{figure}

\subsection{Commodity-specific weighted undirected networks\label{sec_wun_dis}}
We now focus on the disaggregated commodity-specific versions of the weighted undirected ITN, representing the trade of single classes of products. 
We therefore repeat the previous analyses after setting $\mathbf{W}\equiv\mathbf{W}^c$ for various individual commodities $c>0$.
As we did for the binary case \cite{part1}, we show our results for a subset of 6 commodities taken from the top 14 categories, namely the two commodities with the smallest traded volume ($c=93,9$), two ones with intermediate volume ($c=39,90$), the one with the largest volume ($c=84$), plus the aggregation of all the top 14 commodities (similar results hold also for the other commodities). Together with the completely aggregated data ($c=0$) considered above, this dataset consists of 7 networks with increasing trade volume and level of aggregation.

\begin{figure}[t]
\includegraphics[width=.45\textwidth]{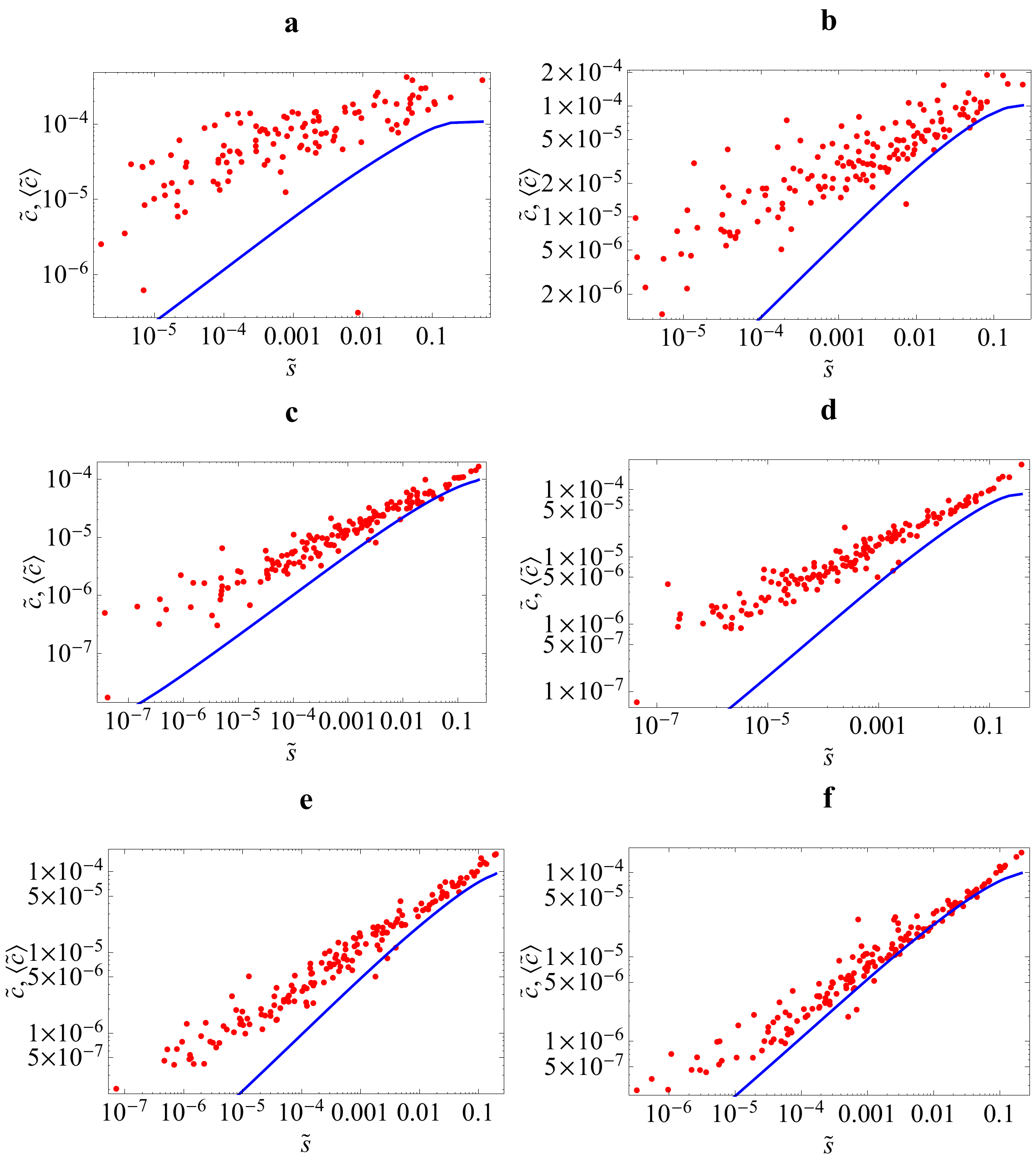}
\caption{\label{fig_wun_dis2}
(Color online) Weighted clustering coefficient $\tilde{c}_i$ versus strength $\tilde{s}_i$ in the 2002 snapshots of the commodity-specific (disaggregated) versions of the real weighted undirected ITN (red points), and corresponding average over the maximum-entropy ensemble with specified strengths (blue solid curves).
\textbf{a)} commodity 93; 
\textbf{b)} commodity 09; 
\textbf{c)} commodity 39; 
\textbf{d)} commodity 90; 
\textbf{e)} commodity 84; 
\textbf{f)} aggregation of the top 14 commodities (see Ref. \cite{part1} for details).
From a) to f), the intensity of trade and level of aggregation increases.} 
\end{figure}

In Fig.~ \ref{fig_wun_dis1}, we show the scatter plot of the average nearest neighbor strength as a function of the strength. Similarly, in Fig.~ \ref{fig_wun_dis2}, we report the scatter plot for the weighted clustering coefficient. Both are shown for the 2002 snapshots of the 6 commodity-specific networks. When compared with the  aggregated network (shown previously in Figs.~\ref{fig_wun_snn} and \ref{fig_wun_cs}), these results lead to interesting conclusions. In general, as happens in the binary case \cite{part1}, we find that commodities with a lower traded volume feature more dispersed scatter plots, with larger fluctuations of the empirical data around the average trend. The effect is more pronounced here than in the binary case. However, while in the latter the real networks are always well reproduced by the null model, in the weighted case the disagreement between empirical and randomized data remains strong across different levels of commodity aggregation. Moreover, the weighted clustering coefficient is the quantity that displays the largest differences between aggregated and disaggregated networks. We see that, for all commodity classes considered, the observed weighted clustering coefficient is  generally larger than its randomized counterpart. However, the deviation is larger for sparser commodities, and decreases as commodity classes with larger trade volumes and higher levels of aggregation are considered. This shows that the partial agreement between real and randomized networks in the completely aggregated case (see Fig.~\ref{fig_wun_cs}) is not robust to disaggregation. In other words, the accordance between empirical data and null model, which according to our discussion in Section \ref{sec_wun_evolution} is  already incomplete in the aggregated case, becomes even worse for sparser commodity-specific networks.

The above results confirm that, unlike the binary case, the properties of the weighted undirected version of the ITN are not completely reproduced by simply controlling for the local properties. The presence of higher-order mechanisms is required as an explanation for the onset and evolution of the observed patterns. This result holds across different years and is enhanced as lower levels of commodity aggregation are considered. 
This shows that a weighted network approach to the analysis of international trade conveys additional information with respect to traditional economic studies that describe trade in terms of local properties alone (total trade, openness, etc.) \cite{Feenstra2004}. Interestingly, a major deviation between the real network and the null model is in the topology implied by local constraints. This confirms, from a different point of view, that in order to properly understand the structure of the international trade system is essential to reproduce its binary topology, even if one is interested in a weighted description.

\section{The ITN as a weighted directed network}
We now turn to the weighted directed analysis of the ITN. A single graph $\mathbf{G}$ in the ensemble of  weighted directed networks is completely specified by its generic weight matrix $\mathbf{W}$ which is in general not symmetric, and whose entry $w_{ij}$ represents the intensity of the directed link from vertex $i$ to vertex $j$ ($w_{ij}=0$ if no directed link is there). 
The binary adjacency $\mathbf{A}$, with entries $a_{ij}=\Theta(w_{ij})$, is in general not symmetric as well.
The \emph{out-strength sequence} $\{s^{out}_i\}$ and the \emph{in-strength sequence} $\{s^{in}_i\}$ represent the local constraints $\{C_a\}$ in the weighted directed case \cite{part1}.
The randomization method \cite{myrandomization} yields the expectation value $\langle X\rangle$ of a property $X$ across the maximally random ensemble of weighted directed graphs with in-strength and out-strength sequences equal to the observed ones (see Appendix \ref{app_wdn}).
The quantities $\{s^{out}_i\}$ and $\{s^{in}_i\}$ (or combinations of them) are now the natural independent variables against which other properties can be visualized in both the real and randomized case, since their expected value coincides with the observed one by construction. 

As for the weighted undirected case, we will consider the rescaled weights $\tilde{w}_{ij}=w_{ij}/w_{tot}$ (where $w_{tot}=\sum_i\sum_{j\ne i}w_{ij}$) in order to wash away trends due to an overall change in the volume of trade across different years.
Correspondingly we consider the rescaled strengths
\begin{eqnarray}
\tilde{s}^{out}_i&\equiv&\sum_{j\ne i}\tilde{w}_{ij}=\frac{s^{out}_i}{w_{tot}}\\
\tilde{s}^{in}_i&\equiv&\sum_{j\ne i}\tilde{w}_{ji}=\frac{s^{in}_i}{w_{tot}}
\end{eqnarray}
and we analogously use $\tilde{w}_{ij}$ instead of $w_{ij}$ in the definition of all quantities.
Note that $w_{tot}=\sum_i s^{in}_i=\sum_i s^{out}_i$, and since $\langle s^{in}_i\rangle=s^{in}_i$ and $\langle s^{out}_i\rangle=s^{out}_i$  we have
\begin{equation}
\langle w_{tot}\rangle=\sum_i \langle s^{in}_i\rangle=\sum_i s^{in}_i=w_{tot}
\label{eq_wtotdir}
\end{equation}
Therefore, as for the undirected case, the expected value of $w_{tot}$ coincides with its empirical value, and the total weight can therefore be safely used to rescale the weights of both real and randomized networks.

\subsection{Directed edge weights}
As we did in Section \ref{sec_wun_w} for the weighted undirected case, we first study the consequences that the specification of the in- and out-strength sequences has on the weights of the network and on its density.

\begin{figure}[t]
\includegraphics[width=.45\textwidth]{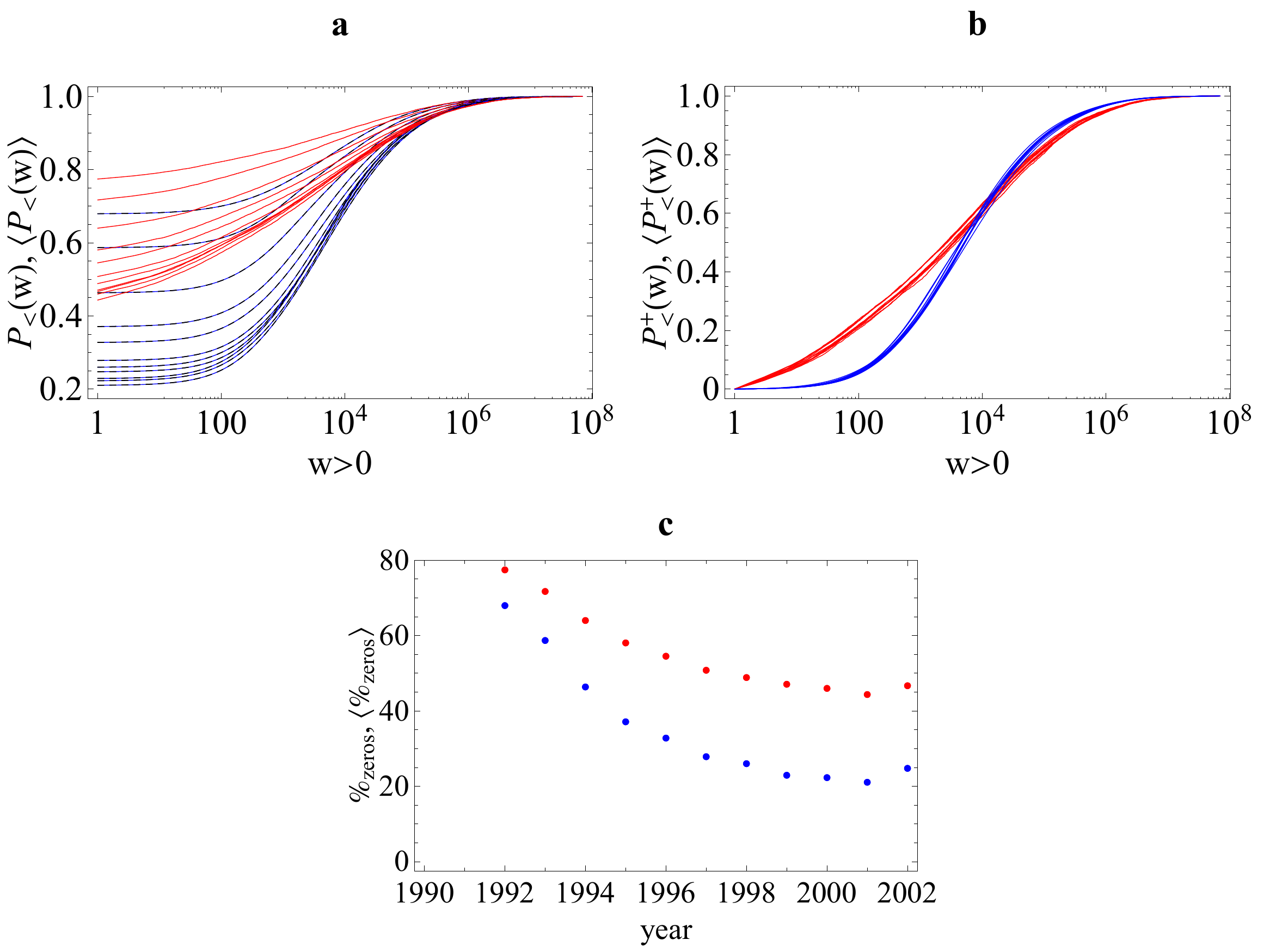}
\caption{\label{fig_wdn_w}
(Color online) Edge weights in the weighted directed ITN. Red (light grey, upper curves/points): real network; blue (dark grey, lower curves/points): expectation for the maximum-entropy ensemble with specified out-strengths and in-strengths.
\textbf{a)} cumulative distributions of edge weights in all years from 1992 (top) to 2002 (bottom). 
\textbf{b)} same as the previous panel, but excluding zero weights (missing links). \textbf{c)} percentage of missing links as a function of time.}
\end{figure}

In Fig.~\ref{fig_wdn_w}a we show the cumulative distribution of observed weights $P_<(w)$ (including missing links with $w=0$) and its randomized counterpart $\langle P_<(w)\rangle$ (see Appendix \ref{app_wdn}). Similarly, in Fig.~\ref{fig_wdn_w}b we show the cumulative distribution of observed positive weights $P^+_<(w)$ (excluding missing links) and the randomized one $\langle P^+_<(w)\rangle$ (see Appendix \ref{app_wdn}).
As in the undirected case, we find that the empirical distributions are always different from the randomized ones, and we confirmed that the hypothesis of equality of real and expected distributions is always rejected by both Kolmogorov-Smirnov and Lilliefors tests ($5\%$ significance level). Similarly, the hypothesis of log-normality of the positive weight distributions  $P^+_<(w)$ and $\langle P^+_<(w)\rangle$ is always rejected ($5\%$ significance level).

In this case too, we can monitor the important difference between the topological density of the real and randomized ITN by plotting the fractions of missing links $\%_{zeros}=P_<(0)$ and $\langle\%_{zeros}\rangle=\langle P_<(0)\rangle$ as a function of time (Fig.~\ref{fig_wdn_w}c). Even if the difference is smaller than in the undirected case, we can confirm on a directed basis that, despite it is usually considered a dense graph, the observed ITN is surprisingly sparser than random directed weighted networks with the same in- and out-strength sequences. Thus the density of (missing) links in the real trade network is not accounted for by size considerations (total imports and total exports of world countries). Again, our use of integer-valued weights ensures that this is not a trivial effect, since the probability $q_{ij}(0)$ of missing links is always positive (see Appendix), and the expected density of links is always strictly smaller than one.

As usual, in what follows we compare higher-order topological properties of the ITN with our null model. We first consider the aggregated snapshot for year 2002 in more detail, then discuss the temporal evolution of the results, and finally perform a study of disaggregated networks.

\subsection{Directed average nearest neighbor strengths}
We consider four generalizations of the definition of the average nearest neighbor strength of a vertex in a directed weighted network:
\begin{eqnarray}
\tilde{s}_{i}^{in/in}&\equiv&\frac
{\sum_{j\ne i}a_{ji}\tilde{s}_{j}^
{in}}{k_{i}^{in}}
=\frac{\sum_{j\ne i}\sum_{k\ne j}a_
{ji}\tilde{w}_{kj}}{\sum_{j\ne i}a_
{ji}}\label{eq_wdn_sinin}\\
\tilde{s}_{i}^{in/out}&\equiv&\frac
{\sum_{j\ne i}a_{ji}\tilde{s}_{j}^
{out}}{k_{i}^{in}}
=\frac{\sum_{j\ne i}\sum_{k\ne j}a_
{ji}\tilde{w}_{jk}}{\sum_{j\ne i}a_
{ji}}\label{eq_wdn_sinout}
\\
\tilde{s}_{i}^{out/in}&\equiv&\frac
{\sum_{j\ne i}a_{ij}\tilde{s}_{j}^
{in}}{k_{i}^{out}}
=\frac{\sum_{j\ne i}\sum_{k\ne j}a_
{ij}\tilde{w}_{kj}}{\sum_{j\ne i}a_
{ij}}\label{eq_wdn_soutin}\\
\tilde{s}_{i}^{out/out}&\equiv&\frac
{\sum_{j\ne i}a_{ij}\tilde{s}_{j}^
{out}}{k_{i}^{out}}
=\frac{\sum_{j\ne i}\sum_{k\ne j}a_
{ij}\tilde{w}_{jk}}{\sum_{j\ne i}a_
{ij}}\label{eq_wdn_soutout}
\end{eqnarray}
Indirect interactions due to chains of
length two (products of the type $a_{ij}
\tilde{w}_{kl}$) contribute to the above
quantities.
A fifth aggregated quantity, which is the
natural analogue of the undirected ANNS,
is based on the (rescaled) total strength
$\tilde{s}^{tot}_i\equiv \tilde{s}^{in}
_i+\tilde{s}^{out}_i$:
\begin{equation}
\tilde{s}_{i}^{tot/tot}\equiv\frac
{\sum_{j\ne i}(a_{ij}+a_{ji})\tilde{s}
_{j}^{tot}}{k_{i}^{tot}}
\label{eq_stottot}
\end{equation}
As in the binary case \cite{part1}, it must be noted that the total (in+out) directed quantities such as $\tilde{s}^{tot}_i$ and $\tilde{s}_{i}^{tot/tot}$ are not trivially related to, and carry more information than, the corresponding undirected  properties $\tilde{s}_i$ and $\tilde{s}^{nn}_i$. In this case, the difference between them is given by the \emph{weighted} reciprocity structure of the network. Unfortunately, there are no well-established measures of reciprocity in the weighted case, and introducing a weighted theory of reciprocity is beyond the scope of the present paper. However, as in the binary case, it is still possible for us to assess, by comparing undirected and total directed weighted properties, whether the reciprocity structure of the directed network changes the results obtained in the undirected case.

\begin{figure}[t]
\includegraphics[width=.4\textwidth]{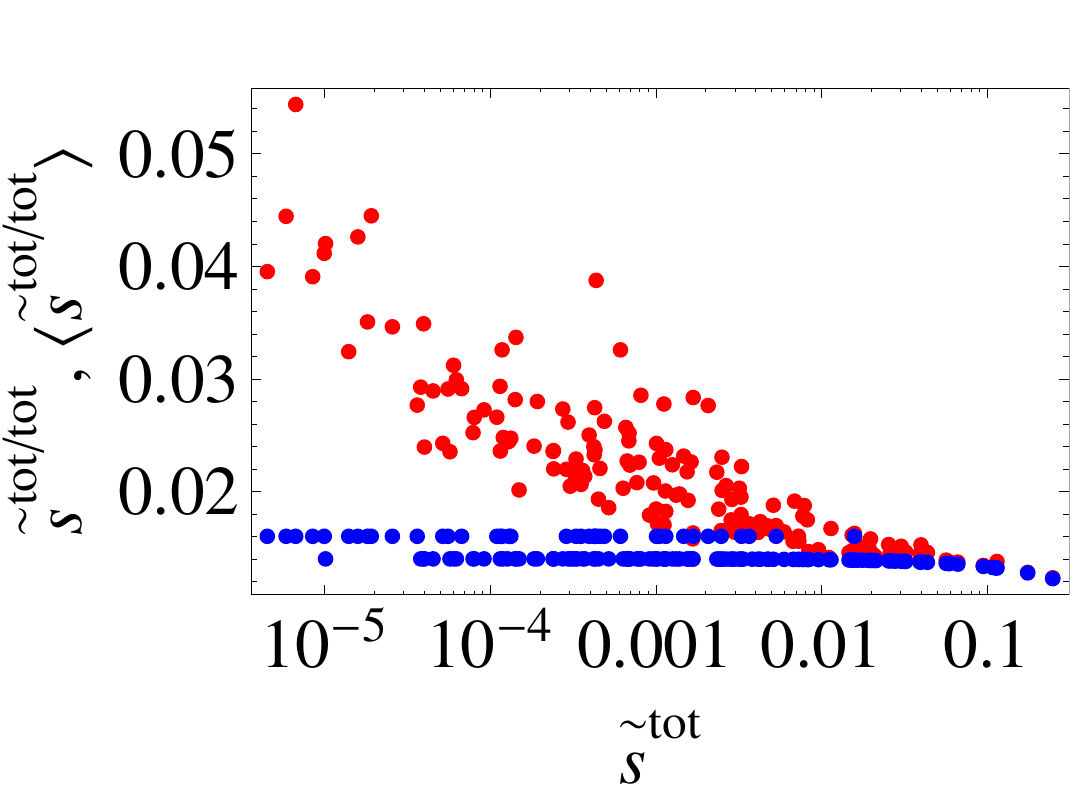}
\caption{\label{fig_wdn_snn} (Color online) Total average nearest neighbor strength $\tilde{s}^{tot/tot}_i$ versus total strength $\tilde{s}^{tot}_i$ in the 2002 snapshot of the real weighted directed ITN (red, upper points), and corresponding average over the null model with specified out-strengths and in-strengths (blue, lower points).}
\end{figure}

In Fig.~\ref{fig_wdn_snn} we show $\tilde
{s}_{i}^{tot/tot}$, together with its randomized value $\langle\tilde
{s}_{i}^{tot/tot}\rangle$ (obtained as in Appendix \ref{app_wdn}), as a function of
$\tilde{s}_{i}^{tot}$ in the aggregated snapshot for year 2002. There are no significant differences with respect to Fig.~\ref{fig_wun_snn}, apart from a ``double'' series of randomized values due to the two possible directions (the terms $a_{ij}$ and $a_{ji}$) that contribute to the definition of $\tilde{s}_{i}^{tot/tot}$ in Eq.~(\ref{eq_stottot}).
Thus we still observe a disassortative behavior in the empirical network, which is not paralleled by the null model.
We now turn to the four directed versions of the ANNS defined in Eqs.(\ref
{eq_wdn_sinin})-(\ref
{eq_wdn_soutout}), as well as their
randomized values (see Appendix \ref{app_wdn}).
As shown in Fig.~\ref{fig_wdn_snndir}, we find that the four empirical quantities all display the same disassortative trend, whereas the four randomized ones are always approximately flat (and no longer switch between two trends as in Fig.~\ref{fig_wdn_snn}). These results show that, as in the undirected representation, the correlation properties of the directed weighted ITN deviate significantly from the ones displayed by the null model with specified strength sequences. In particular, the pronounced disassortativity of the real network is a true signature of negative correlations between the total trade values (in any direction) of neighboring countries, even after controlling for the heterogeneities in the total trade values themselves. This is in marked contrast with the binary case, where we showed that the observed disassortativity is completely explained by controlling for the empirical degree sequence \cite{part1}.
\begin{figure}[b]
\includegraphics[width=.45\textwidth]{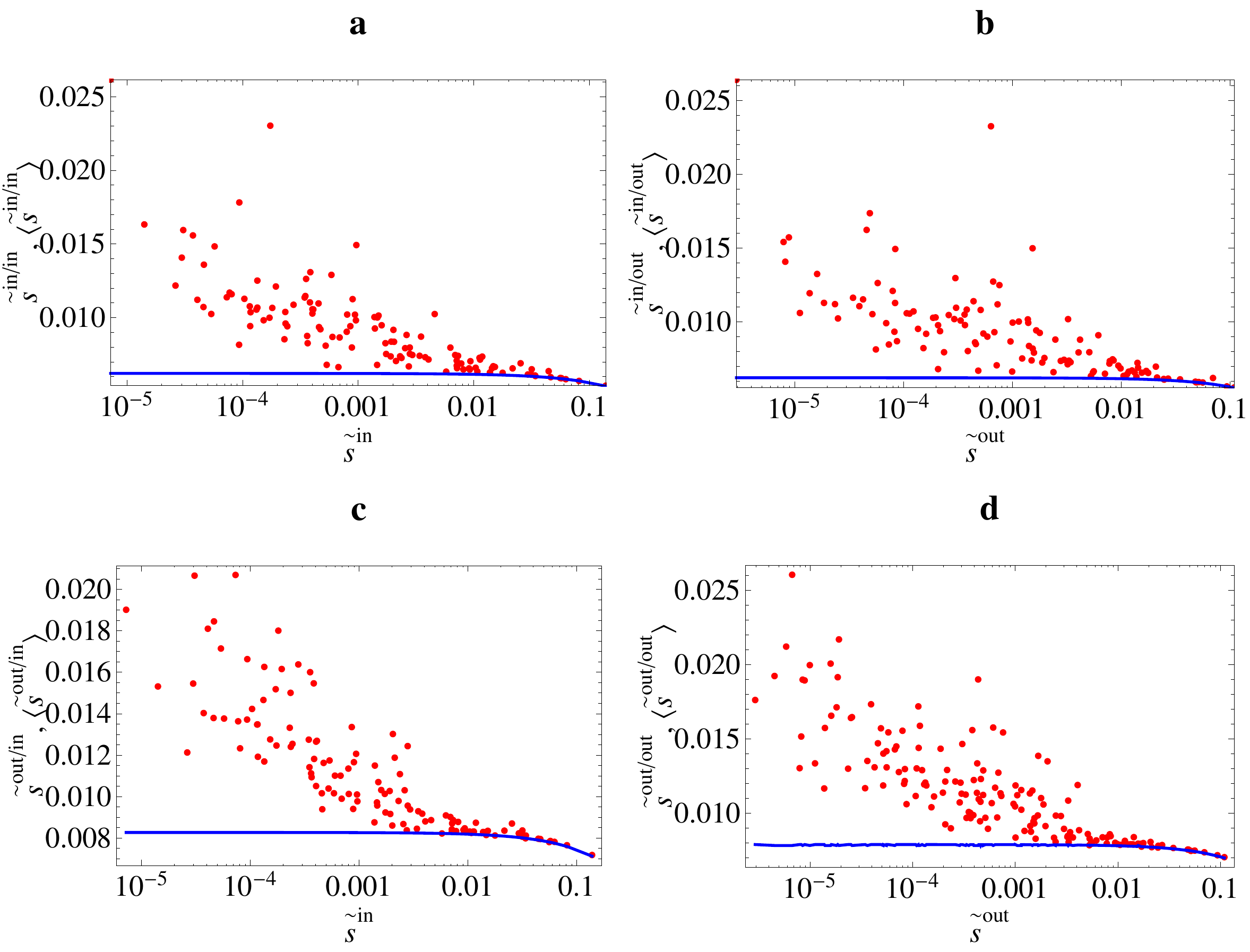}
\caption{\label{fig_wdn_snndir} (Color online) Directed average nearest neighbor strengths versus vertex strengths  in the 2002 snapshot of the real weighted directed ITN (red points), and corresponding averages over the null model with specified out-strengths and in-strengths (blue solid curves). 
\textbf{a)}
$\tilde{s}^{in/in}_i$ versus $\tilde{s}^{in}_i$;
\textbf{b)} $\tilde{s}^{in/out}_i$ versus $\tilde{s}^{in}_i$;
\textbf{c)} $\tilde{s}^{out/in}_i$ versus $\tilde{s}^{out}_i$;
\textbf{d)} $\tilde{s}^{out/out}_i$ versus $\tilde{s}^{out}_i$.}
\end{figure}
\begin{figure}[t]
\includegraphics[width=.45\textwidth]{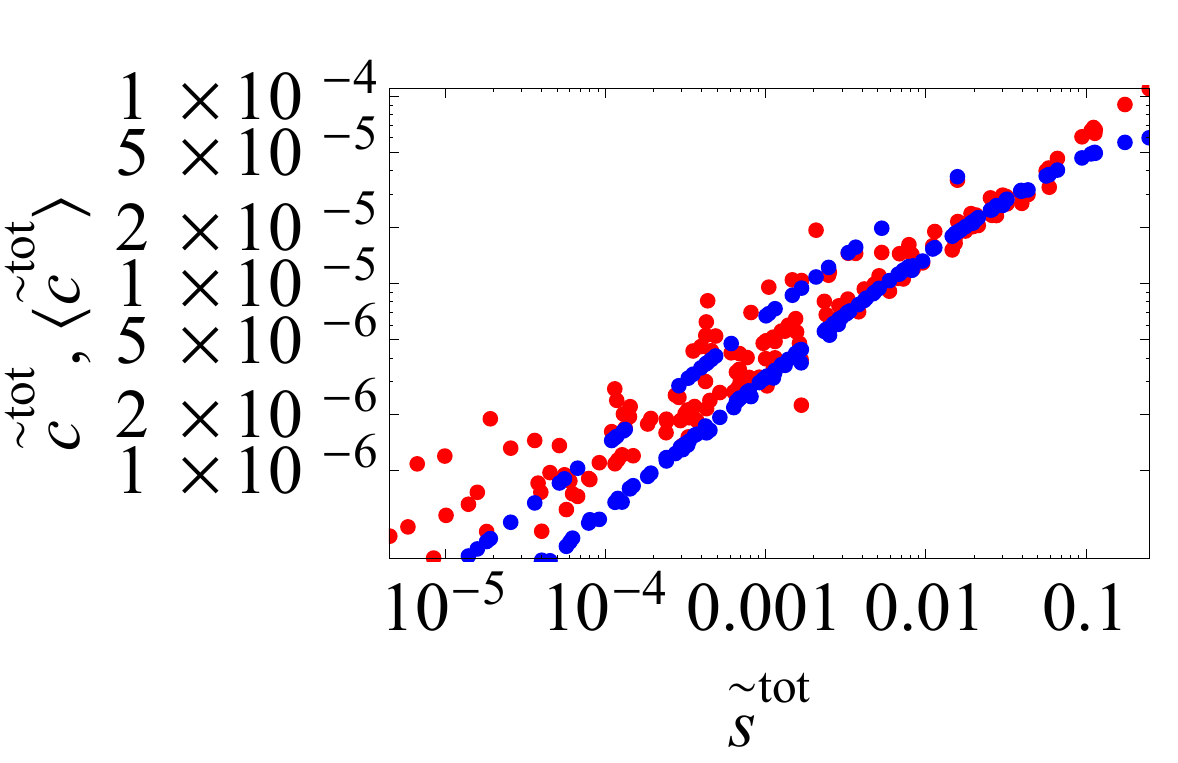}
\caption{\label{fig_wdn_cs} (Color online) Total weighted clustering coefficient $\tilde{c}^{tot}_i$ versus total strength $\tilde{s}^{tot}_i$ in the 2002 snapshot of the real weighted directed ITN (red, lighter points), and corresponding average over the null model with specified out-strengths and in-strengths (blue, darker points).}
\end{figure}
\begin{figure}[b]
\includegraphics[width=.45\textwidth]{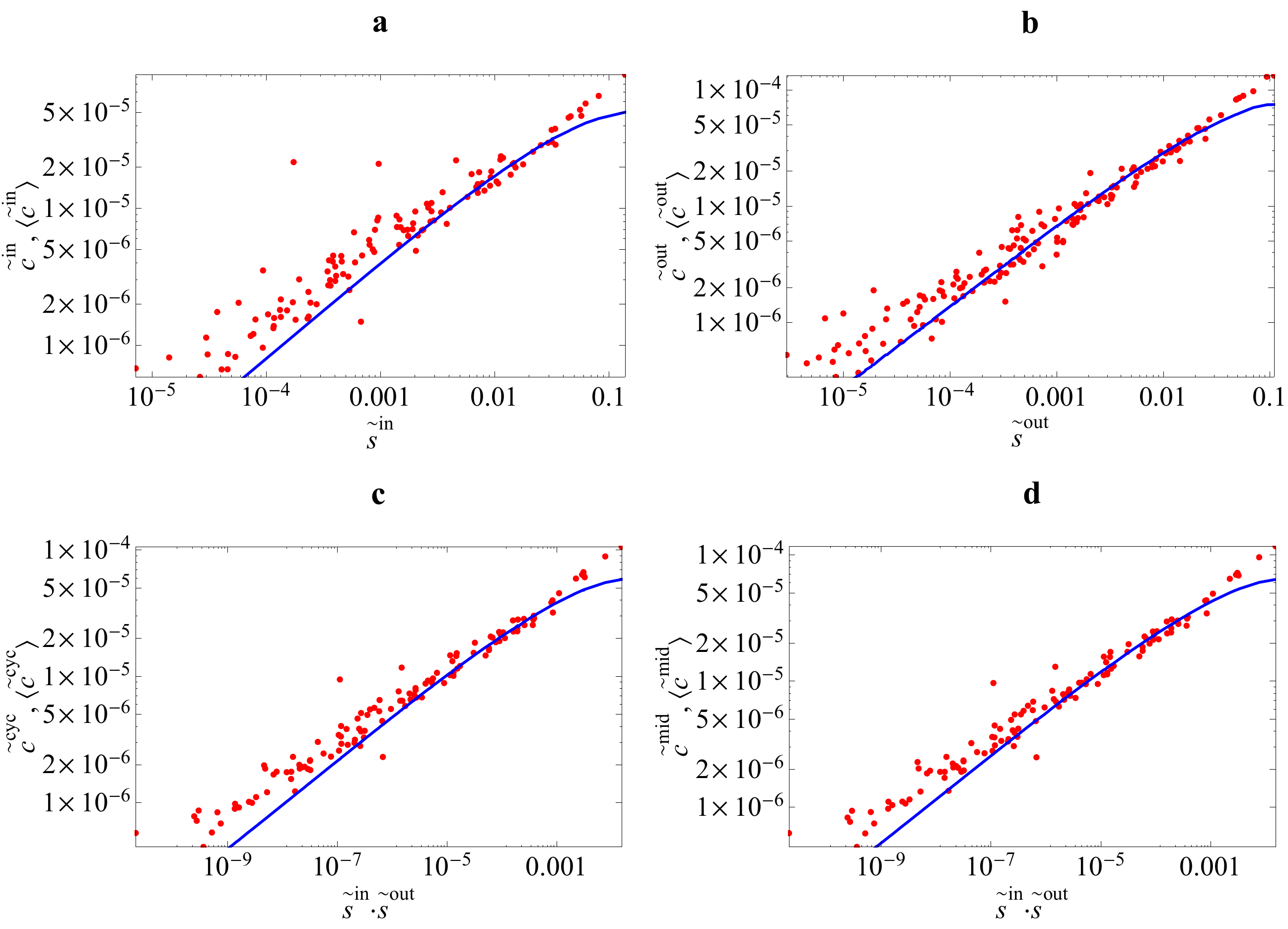}
\caption{\label{fig_wdn_cdir} (Color online) Weighted clustering coefficients versus vertex strengths  in the 2002 snapshot of the real weighted directed ITN (red points), and corresponding averages over the null model with specified out-strengths and in-strengths (blue solid curves).
\textbf{a)} $\tilde{c}^{in}_i$ versus $\tilde{s}^{in}_i$;
\textbf{b)} $\tilde{c}^{out}_i$ versus $\tilde{s}^{out}_i$;
\textbf{c)}  $\tilde{c}^{cyc}_i$ versus $\tilde{s}^{in}_i\cdot\tilde{s}^{out}_i$;
\textbf{d)} $\tilde{c}^{mid}_i$ versus $\tilde{s}^{in}_i\cdot\tilde{s}^{out}_i$.}
\end{figure}
\subsection{Directed weighted clustering coefficients}
In Figs. \ref{fig_wdn_cs} and \ref{fig_wdn_cdir} we report a similar analysis for the clustering coefficient.
The four weighted versions of the \emph
{inward}, \emph{outward}, \emph
{cyclic} and \emph{middleman} directed
clustering coefficients considered in Ref. \cite{part1}
read \cite{Fagiolo2007pre}
\begin{eqnarray}
\tilde{c}_{i}^{in}&\equiv&\frac{\sum_
{j\ne i}\sum_{k\ne i,j}(\tilde{w}_{ki}
\tilde{w}_{ji}\tilde{w}_{jk})^{1/3}}
{k_{i}^{in}(k_{i}^{in}-1)}\label
{eq_wdn_cin}\\
\tilde{c}_{i}^{out}&\equiv&\frac
{\sum_{j\ne i}\sum_{k\ne i,j}(\tilde
{w}_{ik}\tilde{w}_{jk}\tilde{w}_{ij})
^{1/3}}
{k_{i}^{out}(k_{i}^{out}-1)}\label
{eq_wdn_cout}\\
\tilde{c}_{i}^{cyc}&\equiv&\frac
{\sum_{j\ne i}\sum_{k\ne i,j}(\tilde
{w}_{ij}\tilde{w}_{jk}\tilde{w}_{ki})
^{1/3}}
{k_{i}^{in}k_{i}^{out}-k_{i}^
{\leftrightarrow}}\label{eq_wdn_ccyc}\\
\tilde{c}_{i}^{mid}&\equiv&\frac
{\sum_{j\ne i}\sum_{k\ne i,j}(\tilde
{w}_{ik}\tilde{w}_{ji}\tilde{w}_{jk})
^{1/3}}
{k_{i}^{in}k_{i}^{out}-k_{i}^
{\leftrightarrow}}\label{eq_wdn_cmid}
\end{eqnarray}

The above quantities capture indirect interactions of length $3$ according to their directionality, appearing as products of the type $\tilde{w}_{ij}\tilde{w}_{kl}\tilde{w}_{mn}$.
A fifth measure aggregates all
directions:
\begin{eqnarray}
&\tilde{c}_{i}^{tot}\equiv\frac{\sum_{j\ne
i}\sum_{k\ne i,j}(\tilde{w}^{1/3}_{ij}+\tilde
{w}^{1/3}_{ji})(\tilde{w}^{1/3}_{jk}+\tilde{w}^{1/3}_
{kj})(\tilde{w}^{1/3}_{ki}+\tilde{w}^{1/3}_{ik})}
{2\big[k_{i}^{tot}(k_{i}^{tot}-1)-2
k_{i}^{\leftrightarrow}\big]}&\nonumber
\label{eq_wdn_ctot}
\end{eqnarray}

We show the latter in Fig.~\ref{fig_wdn_cs},  and the four directed quantities defined in Eqs.(\ref{eq_wdn_cin})-(\ref{eq_wdn_cmid}) in Fig.~\ref{fig_wdn_cdir}. All properties are shown together with their randomized values (see Appendix \ref{app_wdn}), and plotted against the natural independent variables (or combinations of them).
Again, there is no significant difference with respect to the weighted undirected plot (Fig.~\ref{fig_wun_cs}), apart from the switching behavior of $\langle\tilde{c}^{tot}_i\rangle$ between two trends as already discussed for $\langle\tilde{s}^{tot/tot}_i\rangle$. We find an approximate agreement between real and randomized clustering profiles.

\subsection{Evolution of weighted directed properties}
\begin{figure}[t]
\includegraphics[width=.44\textwidth]{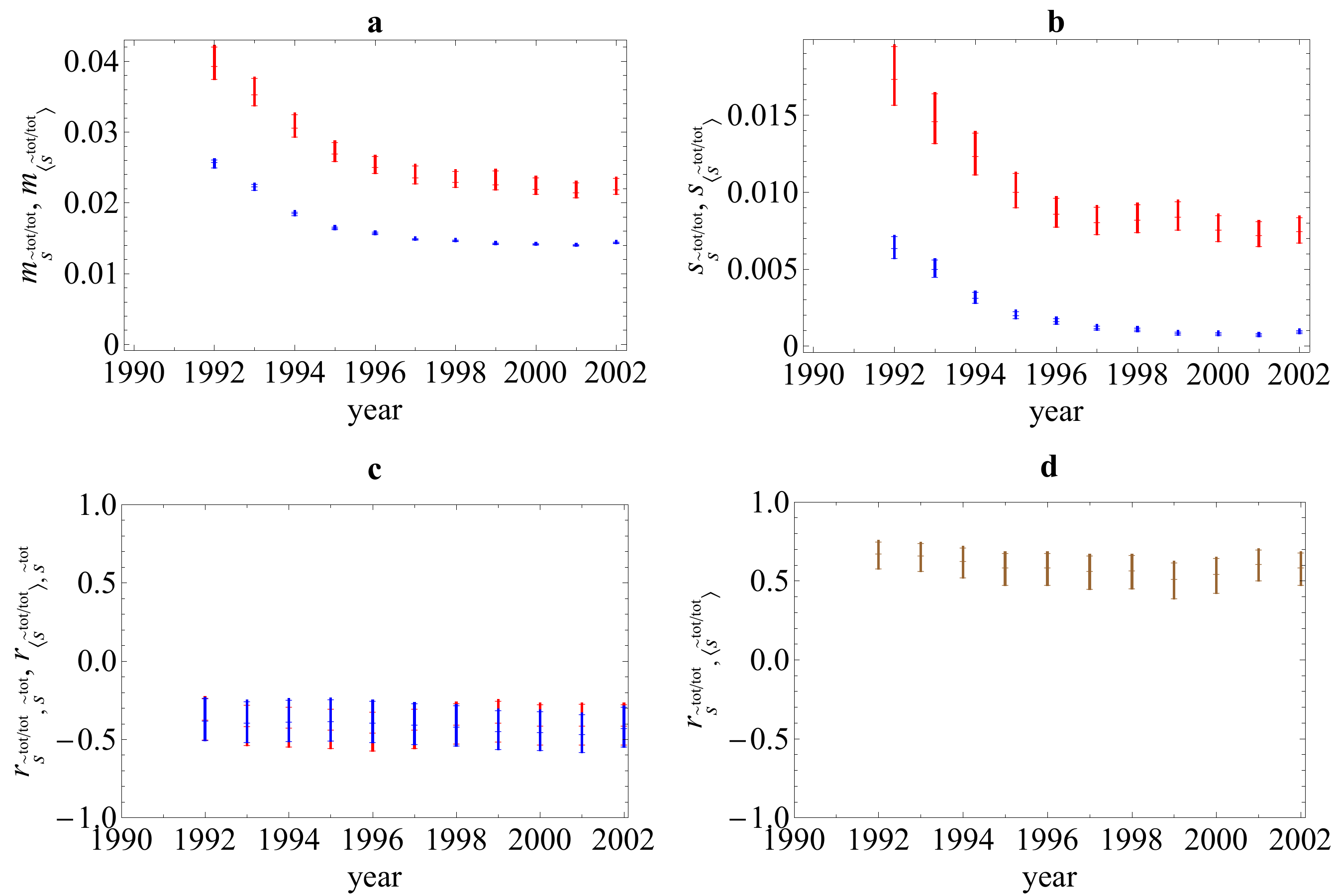}
\caption{\label{fig_wdn_stot_t}
(Color online) Temporal evolution of the properties of the (rescaled) total average nearest neighbor strength $\tilde{s}^{tot/tot}_i$ in the 1992-2002 snapshots of the real weighted directed ITN and of the corresponding null model with specified out-strengths and in-strengths. 
\textbf{a)} average of $\tilde{s}^{tot/tot}_i$ across all vertices (red, upper symbols: real data; blue, lower symbols: null model). 
\textbf{b)} standard deviation of $\tilde{s}^{tot/tot}_i$ across all vertices (red, upper symbols: real data; blue, lower symbols: null model). 
\textbf{c)} correlation coefficient between $\tilde{s}^{tot/tot}_i$ and $\tilde{s}^{tot}_i$ (red: real data; blue: null model, indistinguishable from real data). 
\textbf{d)} correlation coefficient between $\tilde{s}^{tot/tot}_i$ and $\langle \tilde{s}^{tot/tot}_i\rangle$. Vertical bars are $95\%$ confidence intervals.}
\end{figure}

\begin{figure}[b]
\includegraphics[width=.44\textwidth]{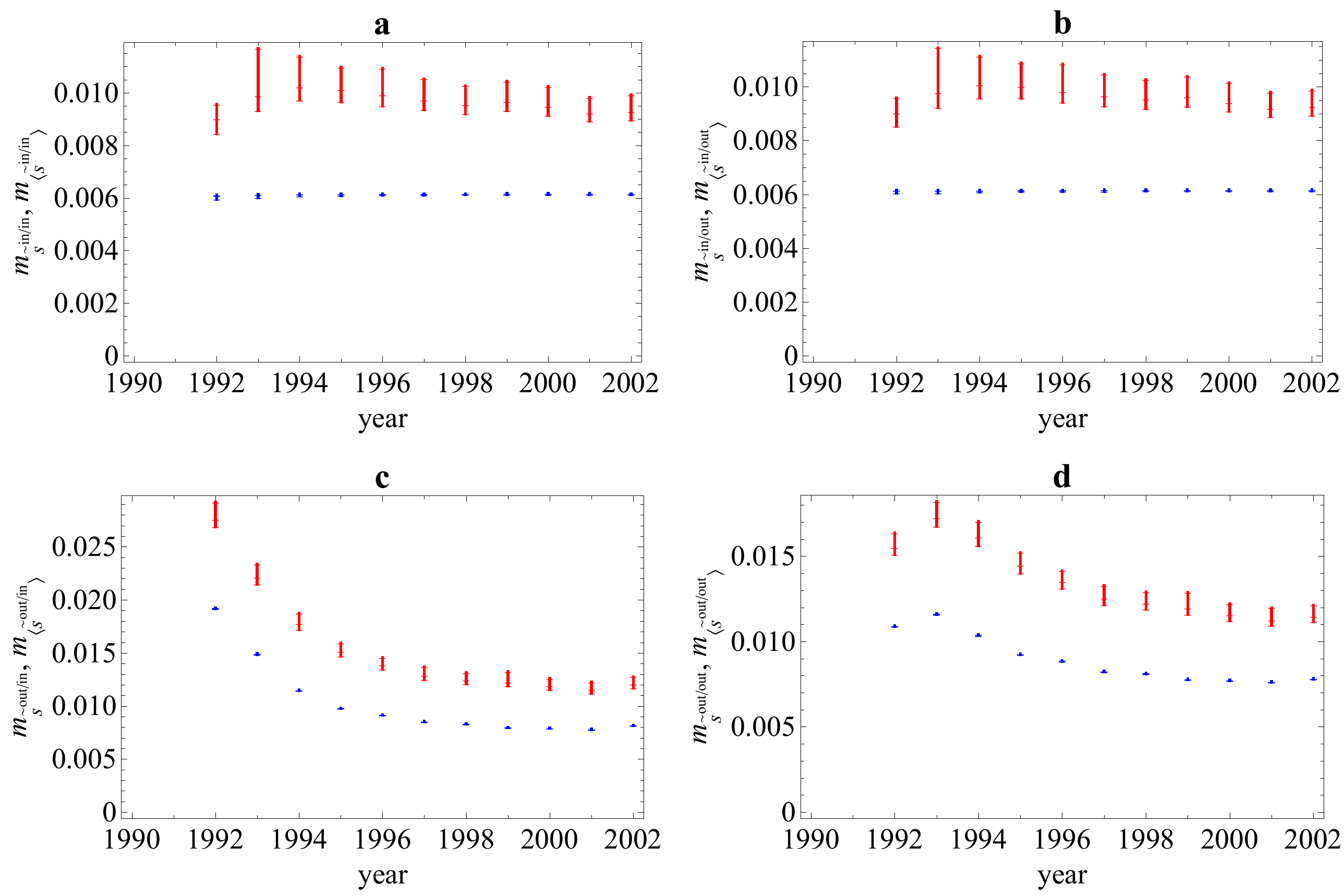}
\caption{\label{fig_wdn_snndir_t} (Color online) Averages and their $95\%$ confidence intervals (across all vertices) of the directed average nearest neighbor strengths in the 1992-2002 snapshots of the real weighted directed ITN (red, upper symbols), and corresponding averages over the maximum-entropy ensemble with specified out-strengths and in-strengths (blue, lower symbols).
\textbf{a)} $\tilde{s}^{in/in}_i$; 
\textbf{b)} $\tilde{s}^{in/out}_i$;
\textbf{c)} $\tilde{s}^{out/in}_i$; 
\textbf{d)} $\tilde{s}^{out/out}_i$.}
\end{figure}

\begin{figure}[t]
\includegraphics[width=.44\textwidth]{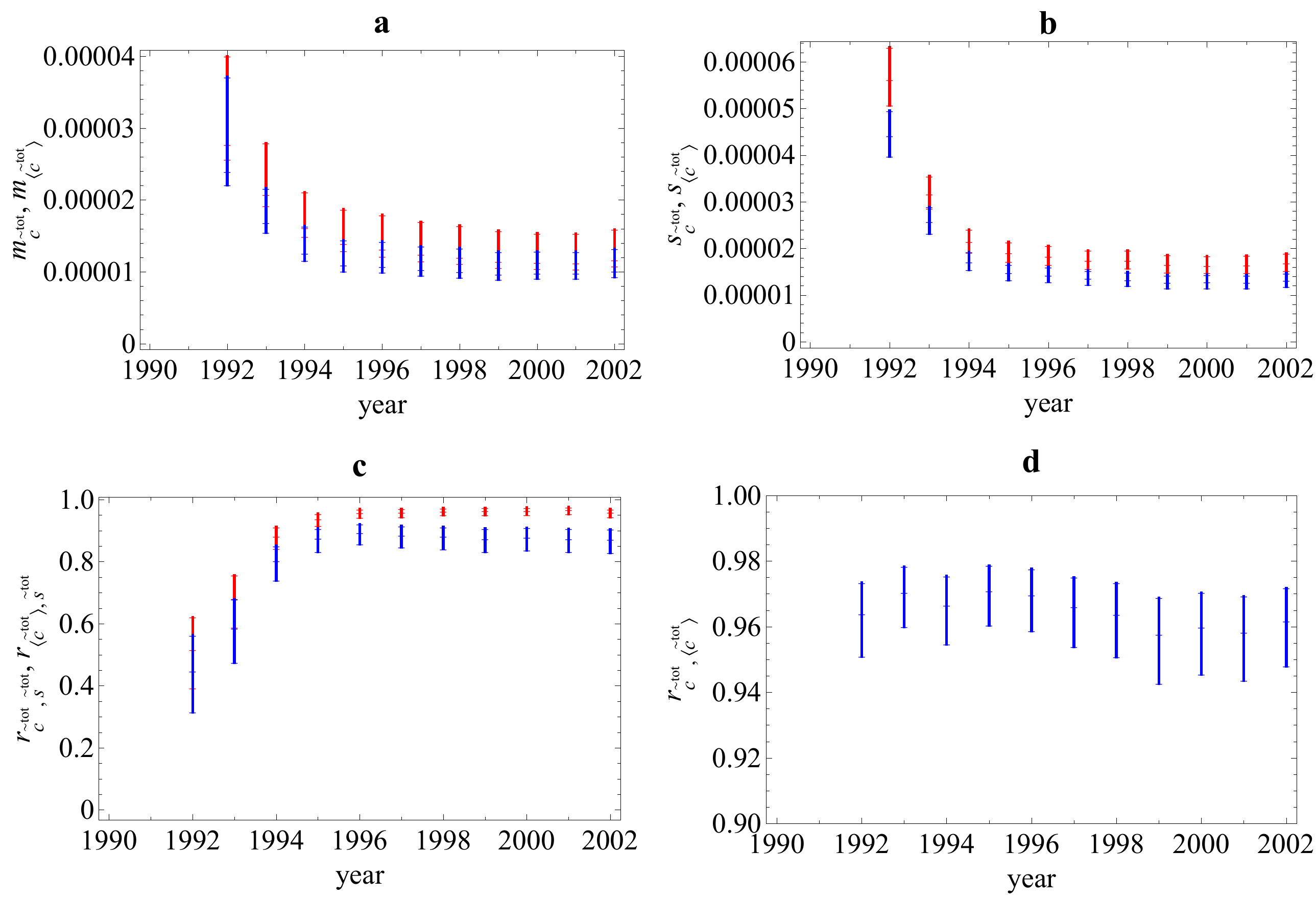}
\caption{\label{fig_wdn_ctot_t}
(Color online) Temporal evolution of the properties of the (rescaled) total weighted clustering coefficient $\tilde{c}^{tot}_i$ in the 1992-2002 snapshots of the real weighted directed ITN and of the corresponding null model with specified out-strengths and in-strengths. 
\textbf{a)} average of $\tilde{c}^{tot}_i$ across all vertices (red, upper symbols: real data; blue, lower symbols: null model). 
\textbf{b)} standard deviation of $\tilde{c}^{tot}_i$ across all vertices (red, upper symbols: real data; blue, lower symbols: null model). 
\textbf{c)} correlation coefficient between $\tilde{c}^{tot}_i$ and $\tilde{s}^{tot}_i$ (red, upper symbols: real data; blue, lower symbols: null model). 
\textbf{d)} correlation coefficient between $\tilde{c}^{tot}_i$ and $\langle \tilde{c}^{tot}_i\rangle$. The $95\%$ confidence intervals of all quantities are represented as vertical bars.}
\end{figure}

\begin{figure}[b]
\includegraphics[width=.44\textwidth]{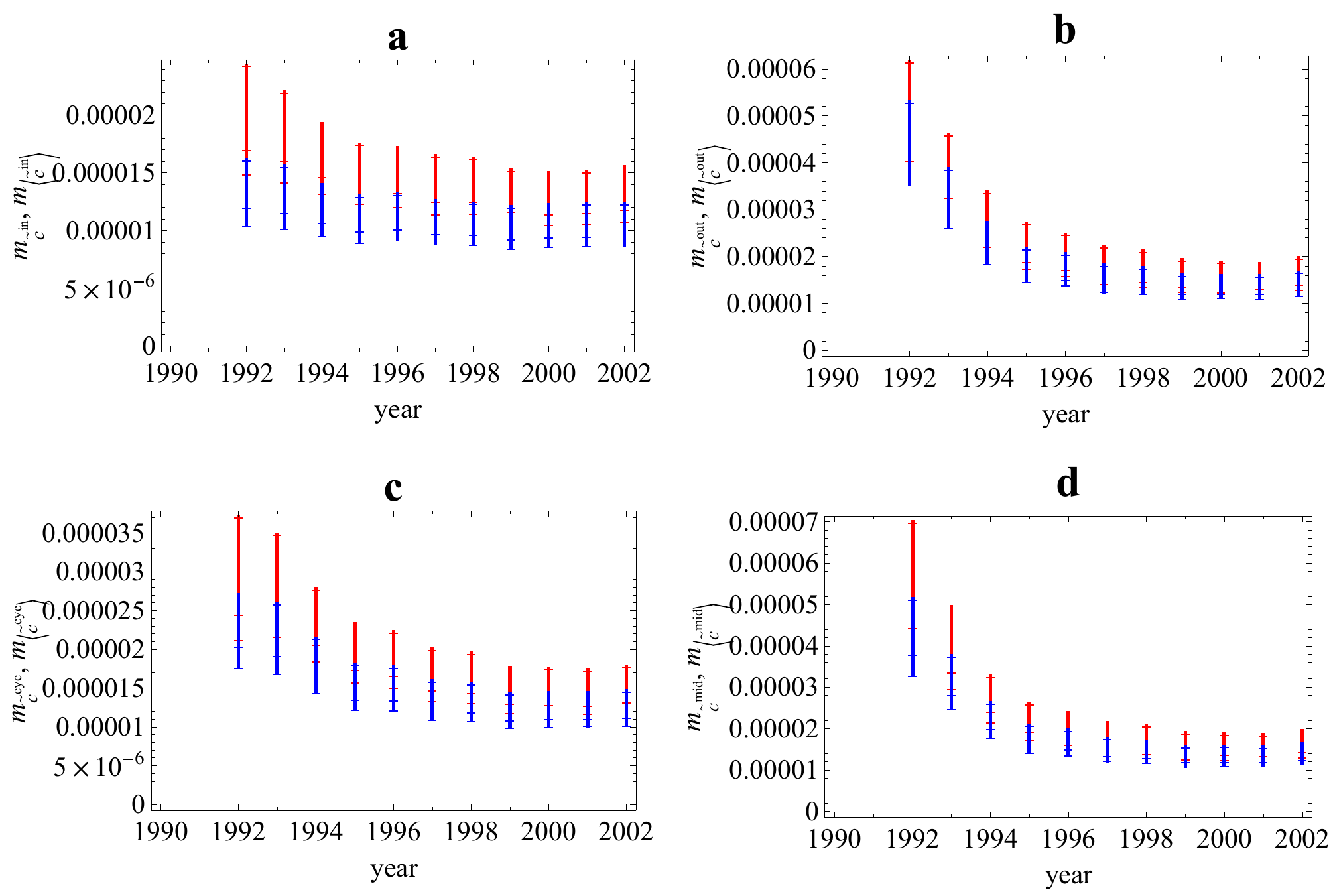}
\caption{\label{fig_wdn_cdir_t} (Color online) Averages and their $95\%$ confidence intervals (across all vertices) of the directed weighted clustering coefficients in the 1992-2002 snapshots of the real weighted directed ITN (red, upper symbols), and corresponding averages over the null model with specified out-strengths and in-strengths (blue, lower symbols).
\textbf{a)} $\tilde{c}^{in}_i$; 
\textbf{b)} $\tilde{c}^{out}_i$;
\textbf{c)} $\tilde{c}^{cyc}_i$;
\textbf{d)} $\tilde{c}^{mid}_i$.}
\end{figure}

In Figs.\ref{fig_wdn_stot_t}-\ref{fig_wdn_cdir_t} we show the temporal evolution of the structural properties considered. Figure \ref{fig_wdn_stot_t} reports the average, standard deviation, and correlation coefficients for $\tilde{s}^{tot/tot}_i$ as a function of time, and Fig.~\ref{fig_wdn_snndir_t} reports (for brevity) only the average of the four directed variants $\tilde{s}^{in/in}_i$, $\tilde{s}^{in/out}_i$, $\tilde{s}^{out/in}_i$, $\tilde{s}^{out/out}_i$.
We find that the detailed description offered by the directed structural properties portrays a different picture with respect to the undirected results shown in Fig.~\ref{fig_wun_snn_t}. In particular, we find that the empirical trends are not always decreasing and the randomized trends are not always constant, in contrast with what previously observed for the undirected ANNS. Both the empirical and randomized values of $\tilde{s}^{tot/tot}_i$ (Fig.~\ref{fig_wdn_stot_t}a) and $\tilde{s}^{out/in}_i$ (Fig.~\ref{fig_wdn_snndir_t}c) display decreasing averages, whereas $\tilde{s}^{in/in}_i$ (Fig.~\ref{fig_wdn_snndir_t}a) and $\tilde{s}^{in/out}_i$ (Fig.~\ref{fig_wdn_snndir_t}b) display constant randomized values and first increasing, then slightly decreasing empirical values. In addition, $\tilde{s}^{out/out}_i$ (Fig.~\ref{fig_wdn_snndir_t}d) displays a different behavior where both real and randomized averages first increase and then decrease. These fine-level differences are all washed away in the undirected description considered in Section \ref{sec_wun}, signaling a loss of information like the one we also observed in the binary case \cite{part1}. However, while in the latter the null model was always in agreement with the empirical data, here we always observe large deviations. In particular, the averages and standard deviations of all empirical quantities are different from their randomized counterparts, and the analysis of the correlation coefficients confirms that the disassortative behavior of the real network is robust in time, and its intensity is systematically not reproduced by the null model.

Different considerations apply to the evolution of the weighted directed clustering coefficients $\tilde{c}^{tot}_i$, $\tilde{c}^{in}_i$, $\tilde{c}^{out}_i$, $\tilde{c}^{cyc}_i$ and $\tilde{c}^{mid}_i$, shown in Figs.~\ref{fig_wdn_ctot_t} and \ref{fig_wdn_cdir_t}.
In this case we find that the undirected trend we observed in Fig.~\ref{fig_wun_c_t} is still not representative of the individual trends of the directed coefficients studied here. However, the empirical and randomized values of the latter are found to be closer here than in the undirected case, and to follow similar temporal behaviors. The null model is however only marginally consistent with the real network, and the knowledge of the strength sequences remains of limited informativeness.

\subsection{Commodity-specific weighted directed networks\label{sec_wdn_dis}}
We finally come to the analysis of disaggregated commodity-specific representations of the weighted directed ITN. We show results for the usual subset of 6 commodity classes ordered by increasing trade intensity ad level of commodity aggregation, to which we can add the completely aggregated case already discussed (again, we found similar results for all commodities).  

\begin{figure}[t]
\includegraphics[width=.45\textwidth]{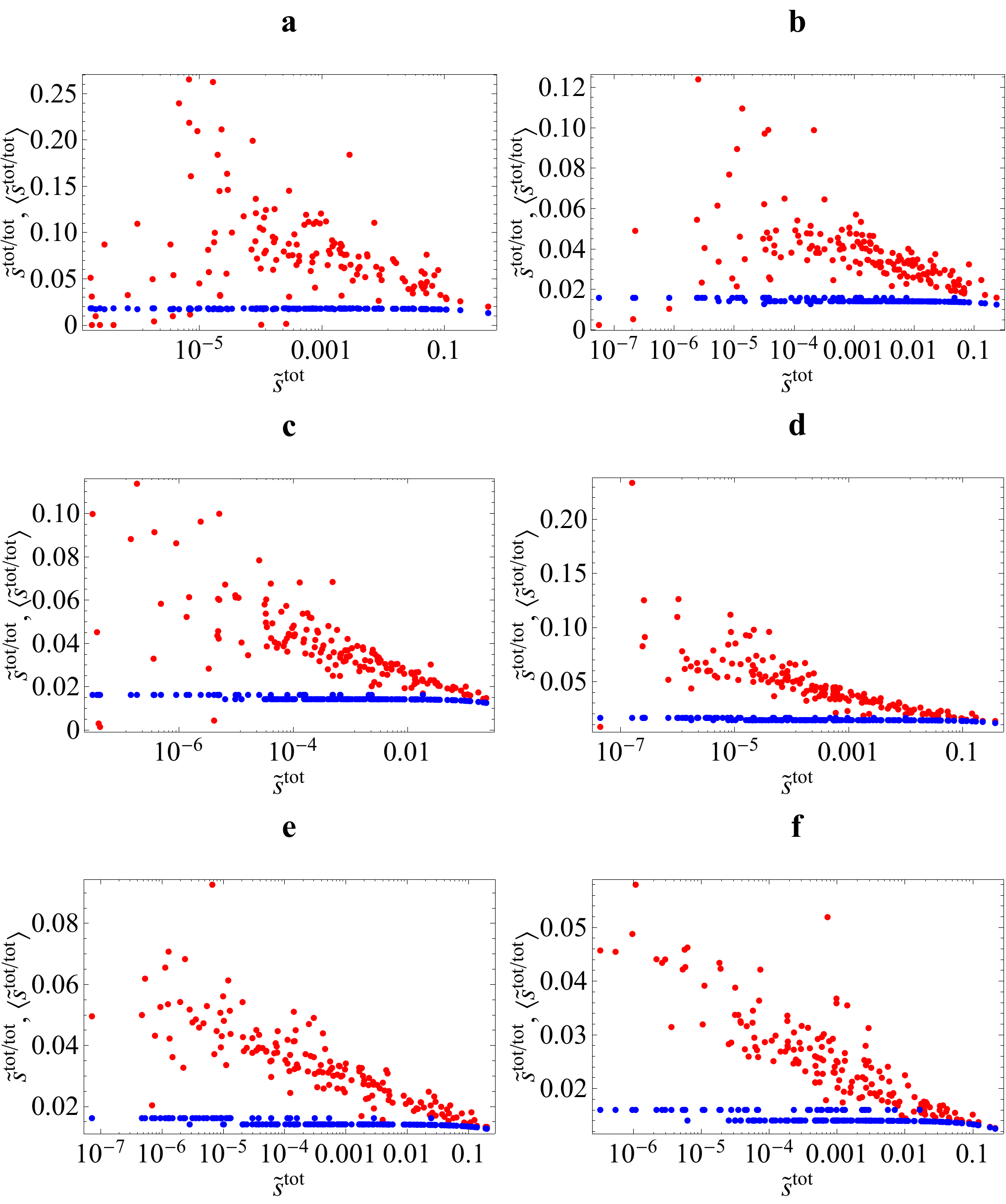}
\caption{\label{fig_wdn_dis1}
(Color online) Total average nearest neighbor strength $\tilde{s}^{tot/tot}_i$ versus total strength $\tilde{s}^{tot}_i$ in the 2002 snapshots of the commodity-specific (disaggregated) versions of the real weighted directed ITN (red, upper points), and corresponding average over the maximum-entropy ensemble with specified out-strengths and in-strengths (blue, lower points).
\textbf{a)} commodity 93; 
\textbf{b)} commodity 09; 
\textbf{c)} commodity 39; 
\textbf{d)} commodity 90; 
\textbf{e)} commodity 84; 
\textbf{f)} aggregation of the top 14 commodities (see Ref. \cite{part1} for details). From a) to f), the intensity of trade and level of aggregation increases.} 
\end{figure}

\begin{figure}[t]
\includegraphics[width=.45\textwidth]{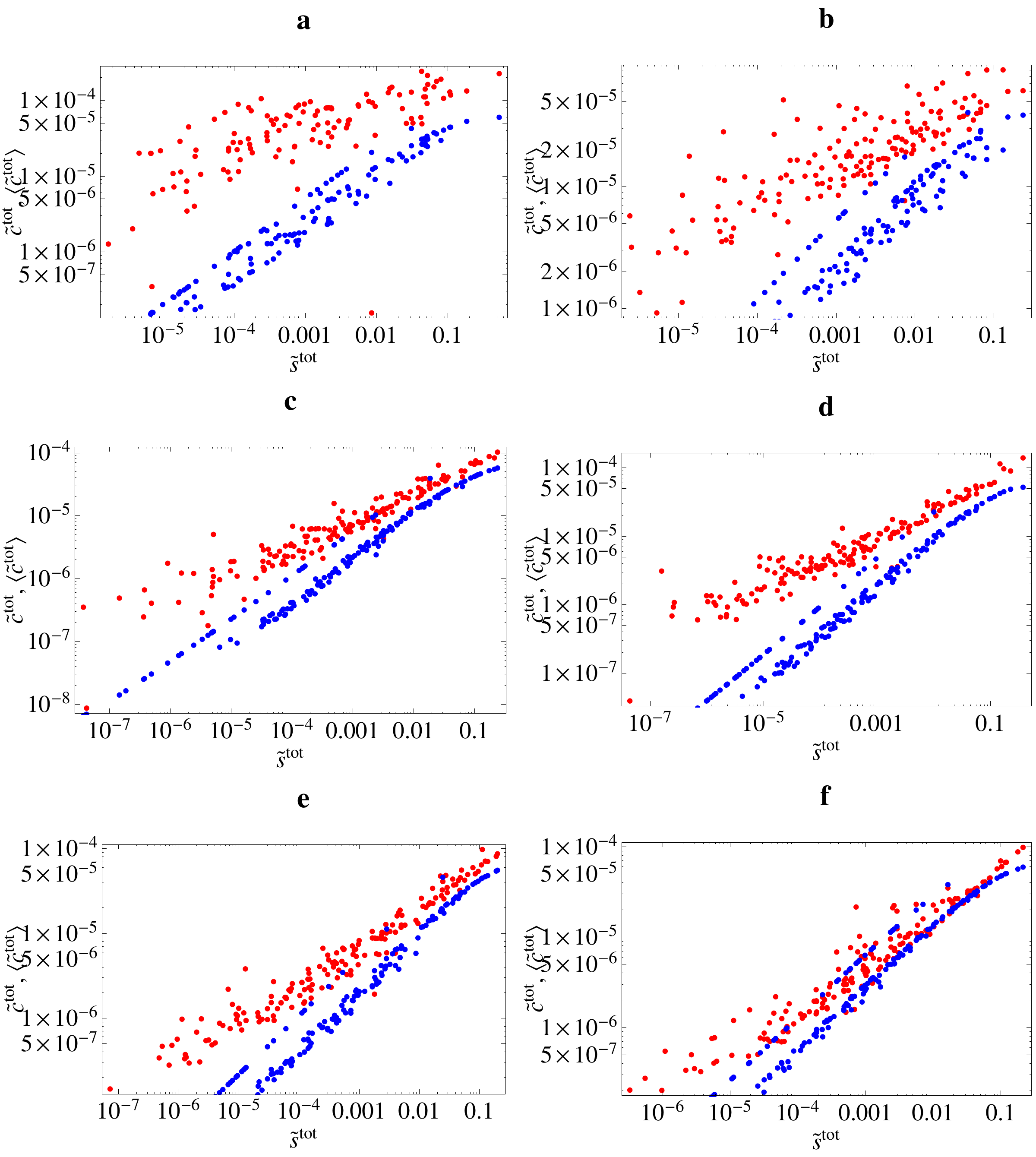}
\caption{\label{fig_wdn_dis2}
(Color online) Total weighted clustering coefficient $\tilde{c}^{tot}_i$ versus total strength $\tilde{s}^{tot}_i$ in the 2002 snapshots of the commodity-specific (disaggregated) versions of the real weighted directed ITN (red, upper points), and corresponding average over the maximum-entropy ensemble with specified our-strengths and in-strengths (blue, lower points).
\textbf{a)} commodity 93; 
\textbf{b)} commodity 09; 
\textbf{c)} commodity 39; 
\textbf{d)} commodity 90; 
\textbf{e)} commodity 84; 
\textbf{f)} aggregation of the top 14 commodities (see Ref. \cite{part1} for details).
From a) to f), the intensity of trade and level of aggregation increases.} 
\end{figure}

Figures \ref{fig_wdn_dis1} and  \ref{fig_wdn_dis2} report the total average nearest neighbor strength and total weighted clustering coefficient as functions of the total strength, for the 6 selected commodity classes in year 2002. 
The corresponding plots for the aggregated networks were shown previously in Figs.~\ref{fig_wdn_snn} and \ref{fig_wdn_cs}.
We find once again that, as more intensely traded commodities and higher levels of aggregation are considered, the empirical data become less scattered around their average trend. In this case, the same effect holds also for the randomized data. 
As for the weighted undirected case, and unlike the binary representation, there is no agreement between empirical networks and the null model. The accordance becomes even worse as commodity classes with smaller trade volume and lower level of aggregation are considered.

The above results extend to the directed case what we found in the analysis of weighted undirected properties. In particular, unlike the binary case, the knowledge of local properties conveys only limited information about the actual structure of the network. Higher-order properties are not explained by local constraints, and indirect interactions cannot be decomposed to direct ones. This holds irrespective of the commodity aggregation level and the particular year considered. This implies that a weighted network approach captures more information than simpler analyses focusing on country-specific local properties. Moreover, simple purely topological properties such as link density are not reproduced by the null model. This implies that, even in weighted analyses, the binary structure is an important property to explain, because it is responsible of major departures of the empirical network from the null model. Therefore, both binary and weighted analyses highlight, for completely different reasons, the importance of reproducing the ITN topology and devoting it more consideration in models of trade.

\section{Conclusions}
In this paper and in the preceding one \cite{part1} we have derived a series of results about the structure of the ITN and the role that local topological properties have in constraining it. Our findings are \emph{a priori} unpredictable without a comparison with a null model, and can be summarized as follows.

In the binary description (both in the directed and undirected cases), we found that specifying the degree sequence(s) (a first-order topological property) is enough to explain higher-order properties \cite{part1}. This result has two consequences. First, it implies that all the observed patterns (disassortativity, clustering, etc.) should not be interpreted as genuine higher-order stylized facts and do not require additional explanations besides those accounting for the different specific numbers of trade partners of all countries. Second, it indicates that the degree sequence encodes virtually all the binary information and is therefore a key structural property that economic models of trade should try to explain in detail.

By contrast, in the weighted description (again, both in the directed and undirected cases) specifying the strength sequence(s) is \emph{not} enough in order to reproduce the other properties of the network. Therefore the knowledge of total trade volumes of all countries is of limited informativeness. A weighted network description of trade, by taking into account indirect interactions besides direct ones, succeeds in conveying additional, nontrivial information with respect to standard economic analyses that explain international trade in terms of local country-specific properties only. In particular, in this case the disassortative character of the network and the high level of clustering cannot be simply traced back to the observed local trade volumes and require additional explanations. Moreover, the purely binary topology of the real trade network is different and sparser (despite the ITN is traditionally considered an unusually dense network) than the one predicted by the null model with the same strength sequence.

Our results bear important consequences for the theory of international trade. The most commonly used modeling framework, i.e. that of gravity models \cite{GravityBook,Fagiolo2010jeic}, relies on the assumption that the intensity of trade between countries $i$ and $j$ depends only on individual properties of $i$ and $j$ (e.g., their GDP) and on additional pairwise quantities relevant to $i$ and $j$ alone (the distance between them plus other factors either favoring or impeding trade). The irreducibility of weighted indirect interactions to direct ones, that we have shown above, implies that even if gravity models succeed in reproducing the magnitude of isolated interactions, they may fail to capture the complexity of longer chains of relationships in the network. And in any case, gravity models generally predict a fully connected network, i.e. no missing links. As we have shown, much of the deviation between real and randomized networks in the weighted case is precisely due to differences in the bare topology. This means that, in order to successfully reproduce the weighted properties of the ITN, it is essential to correctly replicate its binary structure, confirming (from a completely different perspective) the importance of the latter. This explains why in other studies the weighted properties of the aggregated ITN have been replicated by specifying the strength and the degree of all vertices simultaneously \cite{Bhatta2007a}. Even if it is not the focus of the present work, the effects of a simultaneous specification of the strength sequence and of the degree sequence can be studied in more detail applying the same maximum-likelihood method used here \cite{myrandomization} by exploiting the analytical results available for the corresponding maximum-entropy ensemble of weighted graphs \cite{mybosefermi}.

In the light of the above considerations, it is interesting to mention a recent interesting analysis \cite{Fronczak} where the weighted properties of the ITN have been related to the GDP of world countries, in analogy with the similar study carried out by Garlaschelli and Loffredo \cite{Garla2004} in the binary case. In Ref.\cite{Fronczak}, it was shown that the empirical weights of the ITN can be approximately replicated by a model which exploits the GDP of a country to predict its strength. Such a model is the continuous limit of the null model used here, where the fundamental unit of weight approaches zero and weights become real-valued, rather than integer-valued. As we discussed above, such a model predicts fully connected weighted networks and its structure becomes therefore similar to gravity models. Taken together, the results we presented here and in the companion paper \cite{part1} clearly suggest that a satisfactory minimal model relating the properties of the weighted ITN to the GDP of world countries must not only replicate the strengths (and weights) of vertices, but also the degree sequence (and topology), and should therefore be a combination of the models in refs. \cite{Fronczak} and \cite{Garla2004}. Again, this suggests the use of a maximum-entropy ensemble of weighted graphs with fixed strength and degree sequences \cite{mybosefermi}.

In general, our results indicate that theories and models of international trade are incomplete if they only focus on bilateral trade volumes and local weighted properties as in the case of gravity models, and if they do not include the binary topology of the ITN among the main empirical properties to replicate.

\begin{acknowledgments}
D.G. acknowledges financial support from the European Commission 6th FP (Contract CIT3-CT-2005-513396), Project: DIME - Dynamics of Institutions and Markets in Europe. This work was also supported by the Dutch Econophysics Foundation (Stichting Econophysics, Leiden, Netherlands) with funds from Duyfken Trading Knowledge BV, Amsterdam, Netherlands.
\end{acknowledgments}

\appendix

\section{Weighted undirected properties\label{app_wun}}
In the weighted undirected case, each graph $\mathbf{G}$ is completely specified by its (symmetric) non-negative weight matrix $\mathbf{W}$. The entries $w_{ij}$ of this matrix are integer-valued, since so are the trade values we consider \cite{part1}.
The randomization method we are adopting \cite{myrandomization} proceeds by 

\begin{enumerate}
\item \textit{specifying the strength sequence as the constraint:} $\{C_a\}=\{s_i\}$.
The Hamiltonian therefore reads
\begin{equation}
H(\mathbf{W})=\sum_i\theta_i s_i(\mathbf{W})=\sum_i\sum_{j<i}(\theta_i+\theta_j)w_{ij}
\end{equation}
and one can show \cite{mybosefermi} that this allows to write the graph probability as
\begin{equation}
P(\mathbf{W})=\prod_{i}\prod_{j<i} q_{ij}(w_{ij})
\end{equation}
where
\begin{equation}
q_{ij}(w)=(x_i x_j)^w(1-x_i x_j)
\label{eq_wun_qij}
\end{equation}
(with $x_i\equiv e^{-\theta_i}$) is the probability that a link of weight $w$ exists between vertices $i$ and $j$  in the maximum-entropy ensemble of weighted undirected graphs, subject to specifying a given strength sequence as the constraint;

\item \textit{solving the maximum-likelihood equations,} by setting the parameters $\{x_i\}$ to the values that maximize the likelihood $P(\mathbf{W}^*)$ \cite{myrandomization} to obtain the real network. These values can be found as the solution of the following set of $N$ coupled nonlinear equations \cite{mylikelihood}:
\begin{equation}
\langle s_i\rangle=\sum_{j\ne i}\frac{x_i x_j }{1-x_i x_j}
=s_i(\mathbf{W}^*)\qquad\forall i
\label{eq_means}
\end{equation}
where $\{s_i(\mathbf{W}^*)\}$ is the empirical strength sequence of the particular real network $\mathbf{W}^*$. With this choice, Eq.~(\ref{eq_wun_qij}) yields the exact value of the connection probability in the ensemble of randomized weighted networks with the same average strength sequence as the empirical one.

\item \textit{computing the probability coefficients $q_{ij}(w)$}, by inserting the Maximum-Likelihood values $\{x_i\}$ into Eq.~(\ref{eq_wun_qij}) which allows to easily compute the expectation value $\langle X\rangle$ of any topological property $X$ analytically, without generating the randomized networks explicitly \cite{myrandomization}.
Equation (\ref{eq_means}) shows that, by construction, the strengths of all vertices are special local quantities whose expected and empirical values are exactly equal: $\langle s_i\rangle=s_i$. 

\begin{table*}[]
\centering
\begin{tabular}{|c|c|}
\hline
\mbox{\textbf{Empirical undirected properties}} & \mbox{\textbf{Expected undirected properties}}\\
\hline
$w_{ij}$ & $\langle w_{ij}\rangle=\frac{x_{i}x_{j}}{1-x_{i}x_{j}}$\\
\hline
$\tilde{w}_{ij}=\frac{w_{ij}}{w_{tot}}$ & $\langle \tilde{w}_{ij}\rangle=\frac{\langle w_{ij}\rangle}{\langle w_{tot}\rangle}=\frac{\langle w_{ij}\rangle}{w_{tot}}$\\
\hline
$a_{ij}=\Theta(w_{ij})$ & $\langle a_{ij}\rangle=p_{ij}=x_{i}x_{j}$\\
\hline
$\tilde{s}_{i}=\sum_{j\ne i}\tilde{w}_{ij}=\frac{s_i}{w_{tot}}$ & $\langle \tilde{s}_{i}\rangle=\tilde{s}_{i}$\\
\hline
$k_{i}=\sum_{j\ne i}a_{ij}$ & $\langle k_{i}\rangle=\sum_{j\ne i}p_{ij}$\\
\hline
$\tilde{s}_{i}^{nn}=\frac{\sum_{j\ne i}a_{ij}\tilde{s}_{j}}{k_{i}}$ & $\langle \tilde{s}_{i}^{nn}\rangle=\frac{\sum_{j\ne i}p_{ij}\tilde{s}_{j}}{\langle k_{i}\rangle}$\\
\hline
$\tilde{c}_{i}=\frac{\sum_{j\ne i}\sum_{k\ne i,j}\tilde{w}_{ij}^{1/3}\tilde{w}_{jk}^{1/3}\tilde{w}_{ki}^{1/3}}{\sum_{j\ne i}\sum_{k\ne i,j}a_{ij}a_{ik}}$ & $\langle \tilde{c}_{i}\rangle=\frac{\sum_{j\ne i}\sum_{k\ne i,j}\langle \tilde{w}_{ij}^{1/3}\rangle\langle \tilde{w}_{jk}^{1/3}\rangle\langle \tilde{w}_{ki}^{1/3}\rangle}{\sum_{j\ne i}\sum_{k\ne i,j}p_{ij}p_{ik}}$\\
\hline
$P_<(w)$&$\langle P_<(w)\rangle=1-\frac{\sum_i\sum_{j<i}p_{ij}^w}{N(N-1)/2}$\\
\hline
$P^+_<(w)$&$\langle P^+_<(w)\rangle=1-\frac{\sum_i\sum_{j<i}p_{ij}^w}{\sum_i\sum_{j<i}p_{ij}}$\\
\hline
\mbox{\textbf{Empirical directed properties}} & \mbox{\textbf{Expected directed properties}}\\
\hline
$w_{ij}$ & $\langle w_{ij}\rangle=\frac{x_{i}y_{j}}{1-x_{i}y_{j}}$\\
\hline
$\tilde{w}_{ij}=\frac{w_{ij}}{w_{tot}}$ & $\langle \tilde{w}_{ij}\rangle=\frac{\langle w_{ij}\rangle}{\langle w_{tot}\rangle}=\frac{\langle w_{ij}\rangle}{w_{tot}}$\\
\hline
$a_{ij}=\Theta(w_{ij})$ & $\langle a_{ij}\rangle=p_{ij}=x_{i}y_{j}$\\
\hline
$\tilde{s}^{in}_{i}=\sum_{j\ne i}\tilde{w}_{ji}=\frac{s^{in}_i}{w_{tot}}$ & $\langle \tilde{s}^{in}_{i}\rangle=\tilde{s}^{in}_{i}$\\
\hline
$\tilde{s}^{out}_{i}=\sum_{j\ne i}\tilde{w}_{ij}=\frac{s^{out}_i}{w_{tot}}$ & $\langle \tilde{s}^{out}_{i}\rangle=\tilde{s}^{out}_{i}$\\
\hline
$\tilde{s}_{i}^{tot}=\tilde{s}_{i}^{in}+\tilde{s}_{i}^{out}$ & $\langle \tilde{s}_{i}^{tot}\rangle=\langle\tilde{s}_{i}^{in}\rangle
+\langle\tilde{s}_{i}^{out}\rangle=\tilde{s}_{i}^{tot}$\\
\hline
$k^{in}_{i}=\sum_{j\ne i}a_{ji}$ & $\langle k^{in}_{i}\rangle=\sum_{j\ne i}p_{ji}$\\
\hline
$k^{out}_{i}=\sum_{j\ne i}a_{ij}$ & $\langle k^{out}_{i}\rangle=\sum_{j\ne i}p_{ij}$\\
\hline
${k}_{i}^{tot}={k}_{i}^{in}+{k}_{i}^{out}$ & $\langle {k}_{i}^{tot}\rangle=\langle{k}_{i}^{in}\rangle
+\langle{k}_{i}^{out}\rangle={k}_{i}^{tot}$\\
\hline
$k^{\leftrightarrow}_{i}=\sum_{j\ne i}a_{ij}a_{ji}$ & $\langle k^{\leftrightarrow}_{i}\rangle=\sum_{j\ne i}p_{ij}p_{ji}$\\
\hline
$\tilde{s}_{i}^{in/in}=\frac{\sum_{j\ne i}a_{ji}\tilde{s}_{j}^{in}}{k_{i}^{in}}$ & $\langle \tilde{s}_{i}^{in/in}\rangle=\frac{\sum_{j\ne i}p_{ji}\tilde{s}_{j}^{in}}{\langle k_{i}^{in}\rangle}$\\
\hline
$\tilde{s}_{i}^{in/out}=\frac{\sum_{j\ne i}a_{ji}\tilde{s}_{j}^{out}}{k_{i}^{in}}$ & $\langle \tilde{s}_{i}^{in/out}\rangle=\frac{\sum_{j\ne i}p_{ji}\tilde{s}_{j}^{out}}{\langle k_{i}^{in}\rangle}$\\
\hline
$\tilde{s}_{i}^{out/in}=\frac{\sum_{j\ne i}a_{ij}\tilde{s}_{j}^{in}}{k_{i}^{out}}$ & $\langle \tilde{s}_{i}^{out/in}\rangle=\frac{\sum_{j\ne i}p_{ij}\tilde{s}_{j}^{in}}{\langle k_{i}^{out}\rangle}$\\
\hline
$\tilde{s}_{i}^{out/out}=\frac{\sum_{j\ne i}a_{ij}\tilde{s}_{j}^{out}}{k_{i}^{out}}$ & $\langle \tilde{s}_{i}^{out/out}\rangle=\frac{\sum_{j\ne i}p_{ij}\tilde{s}_{j}^{out}}{\langle k_{i}^{out}\rangle}$\\
\hline
$\tilde{s}_{i}^{tot/tot}=\frac{\sum_{j\ne i}(a_{ij}+a_{ji})\tilde{s}_{j}^{tot}}{k_{i}^{tot}}$ & $\langle \tilde{s}_{i}^{tot/tot}\rangle=\frac{\sum_{j\ne i}(p_{ij}+p_{ji})\tilde{s}_{j}^{tot}}{\langle k_{i}^{tot}\rangle}$\\
\hline
$\tilde{c}_{i}^{in}=\frac{\sum_{j\ne i}\sum_{k\ne i,j}\tilde{w}_{jk}^{1/3}\tilde{w}_{ji}^{1/3}\tilde{w}_{ki}^{1/3}}{k_{i}^{in}(k_{i}^{in}-1)}$ & $\langle \tilde{c}_{i}^{in}\rangle=\frac{\sum_{j\ne i}\sum_{k\ne i,j}\langle \tilde{w}_{jk}^{1/3}\rangle \langle \tilde{w}_{ji}^{1/3}\rangle \langle \tilde{w}_{ki}^{1/3}\rangle}{\sum_{j\ne i}\sum_{k\ne i,j}p_{ji}p_{ki}}$\\
\hline
$\tilde{c}_{i}^{out}=\frac{\sum_{j\ne i}\sum_{k\ne i,j}\tilde{w}_{ik}^{1/3}\tilde{w}_{ij}^{1/3}\tilde{w}_{jk}^{1/3}}{k_{i}^{out}(k_{i}^{out}-1)}$ & $\langle \tilde{c}_{i}^{out}\rangle=\frac{\sum_{j\ne i}\sum_{k\ne i,j}\langle \tilde{w}_{ik}^{1/3}\rangle \langle \tilde{w}_{ij}^{1/3}\rangle \langle \tilde{w}_{jk}^{1/3}\rangle}{\sum_{j\ne i}\sum_{k\ne i,j}p_{ij}p_{ik}}$\\
\hline
$\tilde{c}_{i}^{cyc}=\frac{\sum_{j\ne i}\sum_{k\ne i,j}\tilde{w}_{ij}^{1/3}\tilde{w}_{jk}^{1/3}\tilde{w}_{ki}^{1/3}}{k_{i}^{in}k_{i}^{out}-k_{i}^{\leftrightarrow}}$ & $\langle \tilde{c}_{i}^{cyc}\rangle=\frac{\sum_{j\ne i}\sum_{k\ne i,j}\langle \tilde{w}_{ij}^{1/3}\rangle \langle \tilde{w}_{jk}^{1/3}\rangle \langle \tilde{w}_{ki}^{1/3}\rangle}{\langle k_{i}^{in}\rangle \langle k_{i}^{out}\rangle-\sum_{j\ne i}p_{ij}p_{ji}}$\\
\hline
$\tilde{c}_{i}^{mid}=\frac{\sum_{j\ne i}\sum_{k\ne i,j}\tilde{w}_{ik}^{1/3}\tilde{w}_{jk}^{1/3}\tilde{w}_{ji}^{1/3}}{k_{i}^{in}k_{i}^{out}-k_{i}^{\leftrightarrow}}$ & $\langle \tilde{c}_{i}^{mid}\rangle=\frac{\sum_{j\ne i}\sum_{k\ne i,j}\langle \tilde{w}_{ik}^{1/3}\rangle \langle \tilde{w}_{jk}^{1/3}\rangle \langle \tilde{w}_{ji}^{1/3}\rangle}{k_{i}^{in}k_{i}^{out}-\sum_{j\ne i}p_{ij}p_{ji}}$\\
\hline
$\tilde{c}_{i}^{tot}=\frac{\sum_{j\ne i}\sum_{k\ne i,j}(\tilde{w}_{ij}^{1/3}+\tilde{w}_{ji}^{1/3})(\tilde{w}_{jk}^{1/3}+\tilde{w}_{kj}^{1/3})(\tilde{w}_{ki}^{1/3}+\tilde{w}_{ik}^{1/3})}{2\big[k_{i}^{tot}(k_{i}^{tot}-1)-2 k_{i}^{\leftrightarrow}\big]}$ & $\langle \tilde{c}_{i}^{tot}\rangle=\frac{\sum_{j\ne i}\sum_{k\ne i,j}\langle \tilde{w}_{ij}^{1/3}+\tilde{w}_{ji}^{1/3}\rangle \langle \tilde{w}_{jk}^{1/3}+ \tilde{w}_{kj}^{1/3}\rangle \langle \tilde{w}_{ki}^{1/3}+\tilde{w}_{ik}^{1/3}\rangle}{2\big[\sum_{j\ne i}\sum_{k\ne i,j}(p_{ji}p_{ki}+p_{ij}p_{ik})+2(k_{i}^{in}k_{i}^{out})-2\sum_{j\ne i}p_{ij}p_{ji}\big]}$\\
\hline
$P_<(w)$&$\langle P_<(w)\rangle=1-\frac{\sum_i\sum_{j\ne i}p_{ij}^w}{N(N-1)/2}$\\
\hline
$P^+_<(w)$&$\langle P^+_<(w)\rangle=1-\frac{\sum_i\sum_{j\ne i}p_{ij}^w}{\sum_i\sum_{j\ne i}p_{ij}}$\\
\hline
\end{tabular}
\caption{Expressions for the empirical and expected properties in the weighted (undirected and directed) representations of the network.\label{tab_w}}
\end{table*}

\item \textit{computing he expectation values of higher-order topological properties,} as in Table \ref{tab_w}.
The expressions are derived exploiting the fact that $\langle w_{ij}\rangle=\sum_w w q_{ij}(w)=x_i x_j /(1-x_i x_j)$, and that different pairs of vertices are statistically independent, which implies $\langle w_{ij}w_{kl}\rangle=\langle w_{ij}\rangle \langle w_{kl}\rangle$ if $(i-j)$ and $(k-l)$ are distinct pairs of vertices, whereas $\langle w_{ij}w_{kl}\rangle=\langle w_{ij}^2\rangle$ if $(i-j)$ and $(k-l)$ are the same pair of vertices.
Note that we calculate the expected value of the power of the weight between vertices $i$ and $j$ analytically as follows:
\begin{equation}
\langle w_{ij}^\alpha\rangle\equiv\sum_w w^\alpha q_{ij}(w)=(1-x_i x_j)\mbox{Li}_{-\alpha}(x_i x_j)
\end{equation}
where $\mbox{Li}_n(z)$ denotes the Polylogarithm function defined as
\begin{equation}
\mbox{Li}_n(z)\equiv\sum_{l=1}^\infty\frac{z^l}{l^n}
\label{eq_polylog}
\end{equation}
In this paper, we use the above exact expression instead of the approximation $\langle w_{ij}^\alpha\rangle\approx \langle w_{ij}\rangle^\alpha$ suggested in the original paper introducing the method \cite{myrandomization}.
The adjacency matrix representing the existence of a link (irrespective of its intensity) between vertex $i$ and vertex $j$ is derived from the weight matrix by setting $a_{ij}=\Theta(w_{ij})$, where $\Theta(x)=1$ if $x>0$ and $\Theta(x)=0$ otherwise.
The probability that vertices $i$ and $j$ are connected, irrespective of the edge weight, is now $\langle a_{ij}\rangle=p_{ij}\equiv 1-q_{ij}(0)=x_i x_j$.
In analogy with the expectation values of products of weights, we have 
$\langle a_{ij}a_{kl}\rangle=p_{ij}p_{kl}$ if $(i-j)$ and $(k-l)$ are distinct pairs of vertices, whereas $\langle a_{ij}a_{kl}\rangle=\langle a_{ij}^2\rangle=\langle a_{ij}\rangle=p_{ij}$ if $(i-j)$ and $(k-l)$ are the same pair of vertices.
Finally note that we are interested in studying the quantities obtained using the rescaled weights $\tilde{w}_{ij}=w_{ij}/w_{tot}$. This does not introduce complications, since $\langle w_{tot}\rangle=w_{tot}$ as we have shown in Eq.~(\ref{eq_wtotund}). However, the parameters $\{x_i\}$ are computed as in Eq.~(\ref{eq_means}) before rescaling the strengths, since the original integer weights $w_{ij}$ are the actual degrees of freedom.
\end{enumerate}

\section{Weighted directed properties\label{app_wdn}}
In the weighted directed case, the above results can be generalized as follows.
Each graph $\mathbf{G}$ is completely specified by its non-negative (integer-valued) weight matrix $\mathbf{W}$, which now is in general not symmetric.
The maximum-likelihood randomization method \cite{myrandomization} proceeds in this case by
\begin{enumerate}
\item \textit{specifying both the in-strength and the out-strength sequences as the constraints:} $\{C_a\}=\{s^{in}_i,s^{out}_i\}$.
The Hamiltonian takes the form
\begin{equation}
H(\mathbf{W})=\sum_i\left[\theta^{in}_i s^{in}_i(\mathbf{W})+\theta^{out}_i s^{out}_i(\mathbf{W})\right]
\end{equation}
The above choice leads to the graph probability \cite{myrandomization}
\begin{equation}
P(\mathbf{W})=\prod_{i}\prod_{j\ne i} q_{ij}(w_{ij})
\end{equation}
where
\begin{equation}
q_{ij}(w)=(x_i y_j)^w(1-x_i y_j)
\label{eq_wdn_qij}
\end{equation}
(with $x_i\equiv e^{-\theta^{out}_i}$ and $y_i\equiv e^{-\theta^{in}_i}$) is the probability that a link of weight $w$ exists from vertex $i$ to vertex $j$  in the maximum-entropy ensemble of weighted directed graphs with specified in- and out-strength sequences.

\item \textit{solving the maximum-likelihood equations,} 
by setting the parameters $\{x_i\}$ and $\{y_i\}$ are to the values that maximize the likelihood $P(\mathbf{W}^*)$ \cite{myrandomization} to obtain the real network. These values are found as the solution of the following set of $2N$ coupled nonlinear equations \cite{mylikelihood}:
\begin{eqnarray}
\langle s^{out}_i\rangle&=&\sum_{j\ne i}\frac{x_i y_j }{1-x_i y_j}
=s^{out}_i(\mathbf{W}^*)\qquad\forall i\\
\langle s^{in}_i\rangle&=&\sum_{j\ne i}\frac{x_j y_i }{1-x_j y_i}
=s^{in}_i(\mathbf{W}^*)\qquad\forall i
\label{eq_meansinsout}
\end{eqnarray}
where $\{s^{in}_i(\mathbf{W}^*)\}$ and $\{s^{out}_i(\mathbf{W}^*)\}$ are the empirical in- and out-strength sequences of the particular real directed weighted network $\mathbf{W}^*$. Then Eq.~(\ref{eq_wdn_qij}) yields the exact value of the connection probability in the ensemble of randomized directed weighted graphs with the same average strength sequences as the empirical ones.

\item \textit{computing the probability coefficients $q_{ij}(w)$}, by inserting the Maximum-Likelihood values $\{x_i\}$ and $\{y_i\}$ into Eq.~(\ref{eq_wdn_qij}), which allows to obtain the expectation value $\langle X\rangle$ of any topological property $X$ analytically, avoiding the numerical generation of the random ensemble \cite{myrandomization}.
Now, by construction, the in-strengths and out-strengths of all vertices are special local quantities whose expected and empirical values are exactly equal: $\langle s^{in}_i\rangle=s^{in}_i$ and $\langle s^{out}_i\rangle=s^{out}_i$ as shown in Eq.~(\ref{eq_meansinsout}). 

\item \textit{computing the expectation values of higher-order topological properties} as in Table \ref{tab_w}, obtained using the same prescriptions as in the undirected case, with two differences. The first one is that now
\begin{equation}
\langle w_{ij}^\alpha\rangle\equiv\sum_w w^\alpha q_{ij}(w)=(1-x_i y_j)\mbox{Li}_{-\alpha}(x_i y_j)
\end{equation}
where $\mbox{Li}_n(z)$ is still the Polylogarithm function defined in Eq.~(\ref{eq_polylog}). Thus $\langle w_{ij}\rangle=x_i y_j/(1-x_i y_j)$ and $\langle a_{ij}\rangle=p_{ij}\equiv 1-q_{ij}(0)=x_i y_j$, where $a_{ij}=\Theta(w_{ij})$. The expectation values of other powers of the weight change accordingly. Again, these exact expressions replace the approximation prescribed in the paper introducing the method \cite{myrandomization}.
The second one is that, as in the binary directed case, $(i-j)$ and $(j-i)$ are different (and statistically independent) directed pairs of vertices. Therefore $\langle w_{ij}w_{ji}\rangle=\langle w_{ij}\rangle \langle w_{ji}\rangle$ and $\langle a_{ij}a_{ji}\rangle=p_{ij}p_{ji}$.
Again, we have $\langle w_{tot}\rangle=w_{tot}$ as we have shown in Eq.~(\ref{eq_wtotdir}). Therefore we can still easily obtain the quantities built on the rescaled weights $\tilde{w}_{ij}=w_{ij}/w_{tot}$. As for the weighted undirected case, the parameters $\{x_i\}$ and $\{y_i\}$ are however computed using Eq.~(\ref{eq_meansinsout}) before rescaling the strengths, preserving the original integer weights $w_{ij}$ as the actual degrees of freedom.
\end{enumerate}

\bibliographystyle{apsrev}
\bibliography{rewiring_part2}

\end{document}